\documentclass[sigconf,nonacm]{acmart}

\setcopyright{none}
\renewcommand\footnotetextcopyrightpermission[1]{}

\acmConference{}{}{}
\acmBooktitle{}
\acmYear{}
\acmISBN{}

\usepackage{amsmath}
\usepackage{graphicx}
\usepackage{booktabs}
\usepackage{algorithm}
\usepackage{algpseudocode}
\usepackage{xcolor}
\usepackage{multirow}

\newcommand{\tw}{\operatorname{tw}}

\newcommand{\F}{\mathcal{F}}

\newcounter{definition}
\renewcommand{\thedefinition}{\arabic{definition}}
\newcommand{\Def}{%
  \refstepcounter{definition}%
  \noindent\textbf{Definition~\thedefinition:}\ %
}

\newcounter{var}
\renewcommand{\thevar}{\arabic{var}}
\newcommand{\Var}[1]{%
  \refstepcounter{var}%
  \par\medskip\noindent\textbf{Variable~\thevar:}\ \textbf{#1.}\ %
}

\newcounter{encoding}
\renewcommand{\theencoding}{\arabic{encoding}}

\newcommand{\Enc}[1]{%
  \refstepcounter{encoding}%
  \par\medskip\noindent\textbf{Encoding~\theencoding:}\ \textbf{#1.}\ %
}

\makeatletter
\newcommand\paperparagraph{\@startsection{paragraph}{4}{\z@}%
  {0.8ex \@plus 0.2ex \@minus 0.2ex}%
  {-1em}%
  {\normalfont\bfseries}}
\makeatother

\begin{document}
\makeatletter
\renewcommand{\thedefinition}{\arabic{definition}}
\makeatother

\title{SAT-Driven Exact Treewidth via Safe Separators and Structure-Aware Encodings}

\author{Yannik Krone}
\authornote{This work was carried out in the Information and Network Dynamics (INDY) group at EPFL and supervised by Sepehr Elahi. The INDY group is led by Prof.\ Matthias Grossglauser and Prof.\ Patrick Thiran.}
\affiliation{
  \institution{École Polytechnique Fédérale de Lausanne (EPFL)}
  \city{Lausanne}
  \country{Switzerland}
}

\begin{abstract}
Treewidth is a standard measure of graph structure, but computing it exactly is NP-hard. SAT- and MaxSAT-based approaches provide a uniform way to solve treewidth by encoding it in conjunctive normal form and delegating the search to modern Boolean solvers, thus directly benefiting from continued progress in SAT technology. This report studies how to strengthen such SAT-driven solvers with structural reasoning and low-overhead guidance, both at the meta-algorithmic level and within the encoding itself. Starting from a baseline SAT/MaxSAT formulation, we investigate structure-based simplifications that reduce solver effort while preserving exactness, and we evaluate practical heuristic enhancements (fast upper/lower bounds and solver guidance) within iterative SAT and MaxSAT workflows. Experiments show consistent speedups on instances with exploitable structure, and clarify where SAT-based approaches remain limited.
\end{abstract}


\maketitle

\section{Introduction}
Treewidth is a central structural graph parameter that captures how close a graph is to a tree and,
more importantly, when dynamic programming becomes feasible on otherwise intractable problems.
A wide range of NP-hard tasks become tractable on graphs of bounded treewidth because the
combinatorial explosion can be confined to the width of a tree decomposition rather than the total
instance size~\cite{Bodlaender2005DiscoveringTreewidth,Dechter1999BucketElimination}.
The main issue is that computing treewidth exactly is itself NP-hard, so practical treewidth solvers must
balance generality against scalability.
The literature offers a broad range of tools to tackle the problem of treewidth, from elimination-based algorithms and branch-and-bound search
to dynamic programming over decompositions and structural reductions ~\cite{BodlaenderKoster2011LowerBounds,BodlaenderKoster2010UpperBounds, Bodlaender2005DiscoveringTreewidth}.
These approaches are often complementary: heuristics provide quick certificates, while exact
procedures refine them to optimality when needed. \\

\noindent It is also known that treewidth can be characterized via a vertex (elimination) ordering to which a corresponding tree-decomposition can be associated, and that among all vertex orderings the optimal treewidth is attained by at least one ordering~\cite{Dechter1999BucketElimination,Bodlaender2005DiscoveringTreewidth}. Building upon this characterization, SAT- and MaxSAT-based methods offer a uniform, solver-driven route to exact treewidth computation. Instead of engineering dedicated branching rules over elimination orderings, one can encode the
existence of an ordering of width at most $w$ into a conjunctive normal form (CNF) and delegate the search to a modern SAT solver.\\

\noindent Berg and J{\"a}rvisalo~\cite{BergJarvisalo2014SATTreewidth} developed and evaluated an effective
formulation combining an ordering/triangulation encoding with incremental cardinality constraints,
supporting iterative SAT and a MaxSAT optimization variant.
This formulation cleanly separates modeling from search, but raises a natural question:
How can SAT-based treewidth solving be improved in practice by leveraging structural knowledge of the graph more effectively? \\

\noindent This report studies SAT-driven exact treewidth computation through the lens of decomposition.
We take the SAT/MaxSAT framework of Berg and J{\"a}rvisalo as a baseline and investigate how far it
can be pushed by integrating safe-separator reasoning.
Safe separators, introduced by Bodlaender and Koster~\cite{BodlaenderKoster2003SafeSeparators},
support provably sound decomposition rules: splitting a graph along such a separator reduces
treewidth computation to independent subinstances that can be solved separately and recombined
without changing the optimum. We develop two complementary integrations of safe separators.
The first is algorithmic: we wrap the baseline SAT/MaxSAT solver in a recursive procedure that
decomposes along safe separators and invokes SAT/MaxSAT only on residual subgraphs, merging the
resulting certificates into a global tree-decomposition.
The second is encoding-level: we extract a safe-separator decomposition once and flatten it into a
single SAT instance by encoding all leaf subgraphs and linking their
width constraints globally.
These two perspectives expose a trade-off: recursion exploits decomposition opportunities locally,
while flattening reduces repeated solver overhead by letting one incremental solve handle all leaves. Beyond decomposition, we also consider lightweight SAT-layer enhancements that preserve exactness.
First, inexpensive polynomial-time upper and lower bounds restrict the width region that must be
explored, avoiding provably irrelevant SAT/MaxSAT queries in iterative schemes.
Second, heuristic elimination orderings obtained while computing an upper bound can be translated
into guidance for the SAT solver, biasing the search toward solutions that resemble good heuristic
certificates without adding constraints.
We evaluate these enhancements both in isolation and end-to-end in combination with separator-based
reasoning.\\

\noindent Overall, the report makes four contributions.
First, it provides an implementation and empirical baseline for SAT/MaxSAT treewidth solving following the iterative encoding of Berg and J{\"a}rvisalo~\cite{BergJarvisalo2014SATTreewidth}.
Second, it extends this baseline with inexpensive heuristic bounding and low-overhead solver guidance to reduce the number and difficulty of SAT/MaxSAT queries.
Third, it leverages the enhanced SAT/MaxSAT core to exploit separator structure in two complementary ways: a recursive meta-solver that decomposes the instance and solves residual subgraphs exactly, and a separator-aware iterative encoding that compiles the decomposition into a single incremental SAT/MaxSAT instance.
Fourth, it empirically evaluates these techniques, both individually and in combination, across diverse benchmark families, clarifying when they yield substantial gains and when the remaining bottleneck lies in the baseline SAT/MaxSAT encoding.

\section{Preliminaries}
We briefly recall some notions and notation used throughout the report. The definitions in
this section serve as a reference for later sections. Throughout this paper, all graphs are undirected.

\subsection{Treewidth}

\Def Treewidth and tree-decompositions \\
\label{def:treewidth-td}
A tree-decomposition of a graph $G = (V,E)$ is a pair $(T,\{X_t\}_{t \in V(T)})$, where
$T$ is a tree and for each node $t$ of $T$ we have a bag $X_t \subseteq V$, such that:
\begin{enumerate}
  \item Vertex coverage: $\bigcup_{t \in V(T)} X_t = V$.
  \item Edge coverage: For every edge $\{u,v\} \in E$ there is some $t \in V(T)$ with $\{u,v\} \subseteq X_t$.
  \item Running intersection: For every vertex $v \in V$, the set $\{\, t \in V(T) \mid v \in X_t \,\}$ induces a connected subtree of $T$.
    
\end{enumerate}
The width of a tree-decomposition is $\max_{t \in V(T)} |X_t| - 1$.
The \emph{treewidth} of $G$ is the minimum width over all tree-decompositions of $G$, that is,
$\tw(G) = \min_{(T,\{X_t\})} \operatorname{width}(T,\{X_t\})$.

\medskip

\Def Treewidth via linear orderings \\
\label{def:treewidth-elim}
\noindent Fix a linear ordering of the vertices,
$v_1 \prec v_2 \prec \cdots \prec v_n$.
Given an (undirected) edge relation $E' \subseteq \binom{V}{2}$, call $v_i$ a \emph{predecessor} of
$v_j$ (under $\prec$) if $v_i \prec v_j$ and $\{v_i,v_j\}\in E'$, and call $v_j$ a \emph{successor} of
$v_i$ in the same situation.

\medskip\noindent
Starting from $E^{(0)} := E$, repeatedly apply the following rule until saturation:
whenever two vertices $v_i$ and $v_j$ have a common predecessor $v_k$
(i.e., $v_k \prec v_i$ and $v_k \prec v_j$) and both edges $\{v_k,v_i\}$ and $\{v_k,v_j\}$ are present
in the current edge set $E^{(t)}$, add the edge $\{v_i,v_j\}$ if it is not already present.
Let $\Delta(E,\prec)$ be the resulting saturated edge set, and define the induced
triangulation as
\[
  \Delta(G,\prec) \;:=\; (V,\Delta(E,\prec)).
\]

\medskip\noindent
Orient the edges of $\Delta(G,\prec)$ according to $\prec$:
\[
  \tilde{\Delta}(E,\prec)
  \;:=\;
  \{\, (v_i,v_j) \mid \{v_i,v_j\}\in \Delta(E,\prec),\ v_i \prec v_j \,\}.
\]
The \emph{width} of the ordering $\prec$ is the maximum number of successors of any vertex in this
ordered triangulation,
\[
  \operatorname{width}(\prec)
  \;:=\;
  \max_{v_i\in V}\, \bigl|\{\, (v_i,v_j)\in \tilde{\Delta}(E,\prec) \,\}\bigr|.
\]
Optimal treewidth can equivalently be characterized by
\[
  \tw(G) \;=\; \min_{\prec}\ \operatorname{width}(\prec),
\]
that is, $\tw(G)$ is the smallest $w$ for which there exists an ordering $\prec$ with
$\operatorname{width}(\prec)\le w$
\cite{Dechter1999BucketElimination,Bodlaender2005DiscoveringTreewidth,BergJarvisalo2014SATTreewidth}.

\medskip\noindent\textbf{Equivalent elimination and decomposition view. } (Intuition)
The triangulation rule above is equivalent to the standard elimination process:
vertices are eliminated in the order $v_1,v_2,\dots,v_n$; when eliminating $v_i$, all of its
currently remaining neighbors are connected to each other (introducing fill edges), and $v_i$ is
then "removed".
The final saturated graph obtained in this way is exactly $\Delta(G,\prec)$, and the successors of
$v_i$ in $\tilde{\Delta}(E,\prec)$ are precisely the neighbors that are still present when $v_i$ is
eliminated.
Hence, $\operatorname{width}(\prec)$ is the largest neighborhood size that ever needs to be
cliquified. Furthermore, an elimination order yields a tree-decomposition directly.
At the moment $v_i$ is eliminated, form a bag
$X_i := \{v_i\}\cup N_i$, where $N_i$ is the set of its current neighbors.
Then $|X_i|-1 = |N_i|$, so the maximum bag size minus one equals $\operatorname{width}(\prec)$.
By linking each bag $X_i$ to a later bag that contains $N_i$ (via the earliest vertex of
$N_i$ under $\prec$), the bags can be connected into a tree that satisfies the running-intersection
property.
Therefore, minimizing ordering width is equivalent to minimizing tree-decomposition width \cite{Dechter1999BucketElimination,Bodlaender2005DiscoveringTreewidth}.

\subsection{SAT and MaxSAT}

\Def SAT in CNF \\
\label{def:sat}
A CNF formula is a conjunction of clauses, e.g., $F = C_1 \wedge \cdots \wedge C_m$.
Each clause is a disjunction of literals, e.g., $C = (x_1 \vee \neg x_2 \vee x_5)$.
A truth assignment is a map $\pi$ that assigns each Boolean variable $x$ a value in $\{0,1\}$.
The SAT problem asks whether there exists an assignment $\pi$ such that all clauses in $F$ evaluate to true under $\pi$. ~\cite{Biere2009HandbookSAT}\\

\Def MaxSAT \\
\label{def:maxsat}
In (partial) MaxSAT, the clauses are split into hard clauses $F_h$ and soft clauses $F_s$.
The task is to find an assignment $\pi$ that satisfies all clauses in $F_h$ and maximizes the number of satisfied clauses in $F_s$.
In this report, the soft clauses are unweighted (all soft clauses have equal weight). ~\cite{LiManya2009MaxSAT}

\subsection{Safe Separators}

\Def (Safe separator)
\label{def:safe-separator}
A \emph{separator} of a graph $G=(V,E)$ is a vertex set $S\subseteq V$ such that $G\setminus S$
has at least two connected components.
A separator $S$ is \emph{safe for treewidth} if making $S$ a clique does not increase treewidth
\cite{BodlaenderKoster2003SafeSeparators}; writing $K(S)$ as the clique generated by $S$, this means
$\tw(G)=\tw(G\cup K(S))$. \\

\noindent Let $C_1,\dots,C_r$ be the connected components of $G\setminus S$, and define the separator-augmented
subgraphs $G_i' := G[C_i\cup S]\cup K(S)$ for $i=1,\dots,r$.
Then treewidth decomposes as
\begin{equation*}
\label{eq:safe-sep-recombine}
\tw(G) \;=\; \max\Bigl(|S|-1,\ \max_{i=1}^{r}\tw(G_i')\Bigr).
\end{equation*}
Since $K(S)\subseteq G_i'$ implies $\tw(G_i')\ge |S|-1$ for all $i$, an equivalent shorthand is
$\tw(G)=\max_i \tw(G_i')$. We use these two forms interchangeably throughout the report.

\subsection{Lower bounds}
\Def (Minor)
\label{def:minor}
A graph $H$ is a \emph{minor} of a graph $G$ if $H$ can be obtained from $G$ by a sequence of vertex deletions,
edge deletions, and edge contractions.

\noindent Treewidth is \emph{minor-monotone} ~\cite{RobertsonSeymour1986Treewidth, BodlaenderKoster2011LowerBounds} : if $H$ is a minor of $G$, then
\[
  H \text{ is a minor of } G \;\Longrightarrow\; \tw(H) \le \tw(G)
\].

\subsection{Upper bounds}
\Def (Elimination width)
\label{def:elimination-width}
Let $\prec$ be a linear ordering of $V$. Eliminating vertices in this order means that, for each vertex, we make its current neighborhood a clique (add fill edges) and then remove it. The \emph{elimination width} of $\prec$ is
\[
  w(\prec) \;=\; \max_{v \in V} \deg_{\prec}(v),
\]
where $\deg_{\prec}(v)$ denotes the degree of $v$ at the moment it is eliminated. Any ordering $\prec$ yields an upper bound $\tw(G) \le w(\prec)$, and as mentioned, there exists an ordering attaining equality ~\cite{Dechter1999BucketElimination, BodlaenderKoster2010UpperBounds}.  

\section{Baseline: SAT-based Treewidth Computation}
\label{sec:baseline-sat}

This section provides a concise overview of the SAT-based framework for exact treewidth computation introduced by Berg and J{\"a}rvisalo~\cite{BergJarvisalo2014SATTreewidth}, focusing on the components relevant to our subsequent extensions.
SAT concerns the satisfiability of CNF formulas (Definition~\ref{def:sat}).
In this setting, for an undirected graph $G=(V,E)$, Berg and J{\"a}rvisalo encode the decision problem $\tw(G)\le w$ as a CNF formula $\F(G,w)$ such that $\F(G,w)$ is satisfiable
if and only if $G$ admits an elimination ordering of width at most~$w$ (following Definition~\ref{def:treewidth-elim}).

\subsection{Base encoding}
Let $G=(V,E)$ be an undirected graph with $V=\{v_1,\dots,v_N\}$.
The base encoding captures (i) a linear ordering $\prec$ of $V$ and (ii) the ordered triangulation induced by $\prec$. While Berg and J{\"a}rvisalo's encoding  builds on an earlier base encoding~\cite{DalmauKolaitisVardi2002CSPTreewidth}, in this report, we follow their refined variant. We define the following variables:

\Var{Linear Ordering}\label{var:ord}
For every pair $1\le i<j\le N$, introduce a Boolean variable $\mathit{ord}_{ij}$, where $\mathit{ord}_{ij}=1$ encodes $v_i \prec v_j$.
Following Berg and J{\"a}rvisalo~\cite{BergJarvisalo2014SATTreewidth}, we use the derived literal $\mathit{ord}^{*}_{ij}$ defined by
$\mathit{ord}^{*}_{ij}=\mathit{ord}_{ij}$ if $i<j$ and $\mathit{ord}^{*}_{ij}=\neg\mathit{ord}_{ji}$ if $i>j$,
so that $\mathit{ord}^{*}_{ij}=1$ always means $v_i \prec v_j$ for all $i\neq j$.

\Var{Triangulated Graph Edges}\label{var:O}
For all distinct $i,j\in\{1,\dots,N\}$, introduce a Boolean variable $O_{ij}$.
Intuitively, $O_{ij}=1$ means that the ordered triangulated graph induced by $\prec$ contains the directed edge $(v_i,v_j)$,
i.e., $v_j$ is a successor of $v_i$.

\Enc{Base constraints $F_{\text{base}}(G)$}\label{enc:base}
The CNF formula $F_{\text{base}}(G)$ is the conjunction of clause families that enforce the intended meaning of
$\mathit{ord}_{ij}$ and $O_{ij}$.
First, $\mathit{ord}^{*}$ is constrained to form a transitive total order; in particular, for all distinct $i,j,k$,
$(\mathit{ord}^{*}_{ij} \wedge \mathit{ord}^{*}_{jk}) \rightarrow \mathit{ord}^{*}_{ik}$ is enforced.
Second, every original edge must occur in the ordered triangulation in one direction: for each $\{v_i,v_j\}\in E$,
add $O_{ij} \vee O_{ji}$.
Third, fill edges are introduced whenever two vertices share a common predecessor in the ordered triangulation:
for all distinct $i,j,k$, enforce $(O_{ki} \wedge O_{kj}) \rightarrow (O_{ij} \vee O_{ji})$.
Fourth, orientations are made consistent with the order: for all $i\neq j$, enforce $\mathit{ord}^{*}_{ij} \rightarrow \neg O_{ji}$.
Finally, an antisymmetry constraint is added for all $i\neq j$: $\neg O_{ij} \vee \neg O_{ji}$.

\subsection{Full iterative encoding}
\label{subsec:full-iter-encoding}

Recall from Definition~\ref{def:treewidth-elim} that the width of an ordering is the maximum, over all vertices,
of the number of outgoing edges in the ordered triangulated graph. The base CNF $F_{\text{base}}(G)$ already encodes the following constraints: the variables $\mathit{ord}_{ij}$
represent a linear ordering $\prec$, and the variables $O_{ij}$ represent the ordered triangulated graph induced by
$\prec$. Enforcing a width bound $w$ means requiring that no vertex has more than $w$ outgoing $O$-edges in this
ordered triangulation. Using the $O_{ij}$ variables, this becomes a set of cardinality constraints of the form
$C_w(i): \sum_{j\neq i} O_{ij} \le w$ for all $i\in\{1,\dots,N\}$.

\medskip\noindent
\textbf{Using cardinality networks.}
The encoding from Berg and J{\"a}rvisalo    in ~\cite{BergJarvisalo2014SATTreewidth} implements these constraints with cardinality networks ~\cite{AsinEtAl2011CardinalityNetworks,AbioEtAl2013ParametricCardinality},
so that many values of $w$ can be tested incrementally using the same CNF. Fix $i\in\{1,\dots,N\}$ and define the set of out-edge literals $A_i=\{\,O_{ij}\mid j\neq i\,\}$.
A cardinality network encoding $C(A_i)$ introduces auxiliary variables $y^i_1,\dots,y^i_{N-1}$ with the following
threshold meaning: for any assignment $\pi$ and any $k$, if $\pi(y^i_k)=0$ then at most $k-1$ literals in $A_i$ are
true under $\pi$. For example, $\pi(y^i_3)=0$ implies that among the literals in $A_i$
(at most) two are true, i.e., $v_i$ has at most two successors in the ordered triangulation. Therefore, the width constraint $C_w(i)$ is enforced by the single unit clause $\neg y^i_{w+1}$ on top of $C(A_i)$:
if $\pi(\neg y^i_{w+1})=1$, then at most $w$ literals in $A_i$ can be true, which is exactly
$\sum_{j\neq i} O_{ij}\le w$.

\Var{Width-threshold}\label{var:W}
To reuse the same CNF for many values of $w$, introduce variables $W_0,\dots,W_{N-2}$ with
$W_w \leftrightarrow (\neg y^1_{w+1} \wedge \neg y^2_{w+1} \wedge \cdots \wedge \neg y^N_{w+1})$
for each $w\in\{0,\dots,N-2\}$.
Intuitively, $W_w=1$ means that all vertices satisfy the bound “out-degree $\le w$”.
Monotonicity is captured by adding $W_w \rightarrow W_{w+1}$ for all $w\in\{0,\dots,N-3\}$.

\Enc{Berg--J{\"a}rvisalo iterative encoding $F_{\text{iter}}(G)$}\label{enc:bj-iter}
The full iterative CNF is
\begin{equation*}
\makebox[\linewidth][l]{$\displaystyle
\begin{array}{@{}l@{}}
F_{\text{iter}}(G) \;=\;
\underbrace{F_{\text{base}}(G)}_{\text{\scriptsize base (order + triangulation)}}\ \wedge
\\[-0.3mm]
\underbrace{\Bigl(\bigwedge_{i=1}^{N} C(A_i)\Bigr)}_{
  \substack{\text{\scriptsize cardinality networks}\\
            \text{\scriptsize on $A_i=\{O_{ij}\mid j\neq i\}$}}
}\ \wedge\
\underbrace{\Bigl(\bigwedge_{w=0}^{N-2}\bigl(W_w \leftrightarrow \bigwedge_{i=1}^{N}\neg y^i_{w+1}\bigr)\Bigr)}_{
  \substack{\text{\scriptsize width-threshold}}
}\ \wedge\
\underbrace{\Bigl(\bigwedge_{w=0}^{N-3}(W_w \rightarrow W_{w+1})\Bigr)}_{
  \substack{\text{\scriptsize monotonicity}\\
            \text{\scriptsize in $w$}}
}.
\end{array}$}
\end{equation*}

\subsection{Solving for the exact treewidth and tree-decomposition}
\label{subsec:exact-from-iter}

The iterative CNF $F_{\text{iter}}(G)$ (Encoding~\ref{enc:bj-iter}) is built so that different width bounds can be checked
by changing only the assumption on the width-threshold variables $W_w$.
For any $w\in\{0,\dots,N-2\}$, define the width-$w$ decision query as
\[
  F(G,w)\ :=\ F_{\text{iter}}(G)\ \wedge\ W_w .
\]
Equivalently, a SAT solver is called on the fixed CNF $F_{\text{iter}}(G)$ under the single assumption literal $W_w=1$.
By construction, $F(G,w)$ is satisfiable if and only if $\tw(G)\le w$~\cite{BergJarvisalo2014SATTreewidth}.\\

\noindent\textbf{Iterative SAT (search over $w$).}
Since the property $\tw(G)\le w$ is monotone in $w$, Berg and J{\"a}rvisalo~\cite{BergJarvisalo2014SATTreewidth}
recover the exact treewidth by probing values of $w$ until the minimum satisfiable bound is identified.
In linear upward search (\textit{iter-lin-up}), queries start at $w=1$ and increase by one
until the first satisfiable query $F(G,w)$ is found.
In linear downward search (\textit{iter-lin-down}), queries start at $w=N-2$ and decrease
by one while satisfiable; the first unsatisfiable query identifies that the previous width is optimal.
As a corner case, if already $F(G,N-2)$ is unsatisfiable, then no ordering has width at most $N-2$ and
$\tw(G)=N-1$.
In binary search (\textit{iter-bin}), a current interval $[L,U]$ known to contain $\tw(G)$ is
maintained, the midpoint width is tested, and the interval is updated depending on satisfiability. \\

\noindent \textbf{MaxSAT.}
Instead of probing widths iteratively using a SAT solver, the encoding can also be instantiated as a
single partial MaxSAT optimization problem that selects the smallest feasible width
in one call~\cite{BergJarvisalo2014SATTreewidth}.
In MaxSAT optimization (\textit{maxsat}) with equal weights, the instance is
\begin{equation*}
\left(
\underbrace{F_{\text{iter}}(G)}_{\text{\scriptsize hard}},
\ \underbrace{\{\, (W_w,1)\mid w=0,\dots,N-2\,\}}_{\text{\scriptsize soft unit clauses, weight }1}
\right),
\end{equation*}
that is, all clauses of $F_{\text{iter}}(G)$ are hard and each $W_w$ is added as a soft unit clause with
weight~$1$.
Because $W_w \rightarrow W_{w+1}$ holds for all $w$, any assignment satisfying the hard clauses exhibits
a threshold behavior: there exists an index $t$ such that
$W_0,\dots,W_{t-1}$ are false and $W_t,\dots,W_{N-2}$ are true.
Maximizing the number of satisfied soft clauses therefore minimizes this threshold index $t$, which
corresponds exactly to the treewidth $\tw(G)$.

\medskip\noindent
\textbf{Recovering a tree-decomposition.}
Whenever a SAT call returns a satisfying assignment for some optimal width bound $w$ (or MaxSAT returns an
optimal model), we decode the model into an explicit tree-decomposition: we read off an elimination
ordering from the $\textit{ord}_{ij}$ variables and the triangulation (fill) edges from the $O_{ij}$
variables, and convert this information into a tree-decomposition. This decoding step runs in
polynomial time, more details for the implementation is given in
Appendix~\ref{sec:impl-decode-td}.

\medskip
\noindent
\textbf{SAT/MaxSAT backend.}
In the remainder of this report, the solver modes
\textit{iter-lin-up},
\textit{iter-lin-down},
\textit{iter-bin}, and
\textit{maxsat},
all based on the standard Berg--J{\"a}rvisalo encoding,
are collectively referred to as the \emph{(atomic) SAT/MaxSAT backend} or the \emph{regular solver}.
These exact solver modes serve as our baseline and are invoked as a backend inside
\textsc{HybridSolver} (Section~\ref{sec:hybrid}) whenever no further safe-separator decomposition
or additional pruning is available.

\section{Safe-Separator Decomposition}
\label{sec:safe-separators}

We now instantiate the general notion of safe separators (Definition~\ref{def:safe-separator})
with concrete separators.
All separators used in this work are safe, so decomposition preserves optimal treewidth:
after splitting along a safe separator $S$, the resulting subinstances can be solved independently
and recombined. These concrete separators are used later in two places: first as the main decomposition mechanism
inside \textsc{HybridSolver} (Section~\ref{sec:hybrid}), and secondly in our
separator-aware iterative encodings (Section~\ref{sec:sep-aware-encoding}).

\paperparagraph{Disconnected components.}
If $G$ has connected components $G_1,\dots,G_k$, then $S=\varnothing$ is a safe separator and
Definition~\ref{def:safe-separator} yields
\[
  \tw(G) \;=\; \max_{j\in\{1,\dots,k\}} \tw(G_j).
\]
Hence, treewidth can be computed independently on each connected component and combined by taking the maximum. See Section~\ref{sec:impl-disconnected} for more details.

\paperparagraph{Articulation points.}
An articulation point is a vertex $v$ whose removal disconnects the graph, i.e., $S=\{v\}$ is a separator.
Since singleton sets are cliques, turning $S$ into a clique does not change anything, and $S$ is safe by
Definition~\ref{def:safe-separator}. Implementation details on how to detect articulation points are in Section~\ref{sec:impl-articulation}.

\paperparagraph{Clique separators.}
If a separator $S$ induces a clique, then $G\cup K(S)=G$, so $S$ is safe by
Definition~\ref{def:safe-separator}.
In our implementation, clique separators are detected by enumerating cliques up to a small fixed
size bound and checking whether $G\setminus S$ disconnects the graph; see Section~\ref{sec:impl-clique}.

\paperparagraph{Almost-clique separators.}
Almost-clique separators extend clique separators by allowing $S$ to become a clique after removing a a vertex. Under the sufficient conditions identified by Bodlaender and Koster~\cite{BodlaenderKoster2003SafeSeparators},
minimal almost-clique separators are safe for treewidth, so they can be used with the same decomposition-and-recombine scheme
as above. In practice, we generate candidate separators using Tamaki’s heuristic for listing almost-clique minimal
separators~\cite{Tamaki2021AlmostClique}, and then enforce the required safety conditions.

\section{HybridSolver}
\label{sec:hybrid}

This section introduces our \textsc{HybridSolver} (also later referred to as \textit{hybrid}), an exact meta-solver that integrates the safe-separator
machinery of Section~\ref{sec:safe-separators} with the SAT/MaxSAT backends from
Section~\ref{sec:baseline-sat}.
The construction is presented in modular building blocks, which are then combined into the final
recursive solver.

\subsection{Goal and interface}
Given a graph $G=(V,E)$, \textsc{HybridSolver} returns the exact treewidth $\tw(G)$ together with a
tree-decomposition witnessing this width.
Our goal is to spend as much time as possible in inexpensive (polynomial-time) structure and bounds,
and to call the atomic SAT/MaxSAT backend only on residual subinstances where decomposition cannot
proceed further.

\subsection{Building block I: heuristic bounds}
\label{subsec:hybrid-bounds}
At every recursion node, on a subgraph $H=(V_H,E_H)$, we compute a lower bound $LB(H)$ and an upper
bound $UB(H)$ of the treewidth.

\paperparagraph{Lower bound $LB(H)$.}
We compute the lower bound using the Minor-Min-Width  (MMW) heuristic (see Algorithm~\ref{alg:mmw} for more details) from Bodlaender and Koster ~\cite{BodlaenderKoster2011LowerBounds}  which
repeatedly contracts edges to generate a sequence of minors (Definition~\ref{def:minor}) and updates
the bound with the degree of the contracted vertex.
By minor-monotonicity, this yields $LB(H)\le \tw(H)$. The procedure is polynomial-time.

\paperparagraph{Upper bound $UB(H)$.}
We compute the upper bound from a good candidate (greedy) elimination ordering $\prec$ (Algorithm~\ref{alg:greedy-ub}), following Bodlaender and Koster \cite{BodlaenderKoster2010UpperBounds}.
Eliminating vertices along $\prec$ yields an elimination width $w(\prec)$,
and therefore $\tw(H)\le UB(H):=w(\prec)$ (Definition~\ref{def:elimination-width}).
We also convert the elimination trace into a tree-decomposition $\mathcal{D}_{UB}$ of width $UB(H)$
using the standard ordering-to-decomposition construction (See implementation details \ref{sec:impl-decode-td}).
This procedure runs in polynomial time.

\paperparagraph{Immediate pruning when bounds meet.}
If $LB(H)=UB(H)$, then $\tw(H)$ is certified and the recursion terminates on $H$ without invoking SAT/MaxSAT.
This is the cheapest (and often most effective) early-exit condition. We then return the tree-decomposition stemming from the upper-bound.

\subsection{Building block II: safe-separators}
If bounds do not meet, we attempt to split the instance using safe separators from Section~\ref{sec:safe-separators}.

\paperparagraph{Recursion on a safe separator.}
Let $S$ be a found safe separator and let $C_1,\dots,C_r$ be the connected components of $H\setminus S$.
We recurse on the separator-augmented subgraphs
$H_i := H[C_i\cup S]\cup K(S)$, and recombine by Definition~\ref{def:safe-separator}:
\[
  \tw(H) \;=\; \max\Bigl(|S|-1,\max_{i=1}^r \tw(H_i)\Bigr).
\]
We recursively merge returned tree-decompositions through a hub bag containing $S$
(Section~\ref{sec:impl-merge}).

\paperparagraph{Heuristic to pick separators.}
When multiple candidate separators exist (notably for clique / almost-clique enumeration),
we propose to rank candidates using the scoring heuristic of Section~\ref{sec:impl-sep-score},
favoring separators that minimize the size of the largest resulting component.

\subsection{Building block III: bound propagation through decomposition}
\label{subsec:hybrid-bound-prop}
HybridSolver maintains bounds consistently across recursion, so that solving the subinstances
yields a tighter interval for the parent. Accordingly, once all children $H_i$ of a safe-separator
split with separator $S$ have been processed, we update the bounds of $H$ by
\[
\begin{aligned}
LB(H)\;&\gets\;\max\Bigl(LB(H),\ |S|-1,\ \max_i LB(H_i)\Bigr),\\
UB(H)\;&\gets\;\min\Bigl(UB(H),\ \max\bigl(|S|-1,\max_i UB(H_i)\bigr)\Bigr).
\end{aligned}
\]

\subsection{Building block IV: SAT/MaxSAT backend with optimizations}
\label{subsec:sat-enhancement-building-block}

If no safe separator is found, we fall back to the SAT/MaxSAT backend
(Section~\ref{sec:baseline-sat}).
Crucially, it does not call the backend ``cold'': it reuses the heuristic information computed at the
current recursion node---a certified interval $[LB(H),UB(H)]$ and a good elimination ordering---to
guide the exact search without compromising correctness.

\paperparagraph{SAT/MaxSAT backend interface.}
We remind that, on a subinstance $H$, the backend relies on the iterative CNF $F_{\text{iter}}(H)$
(Encoding~\ref{enc:bj-iter}) built from ordering variables $\mathit{ord}_{ij}$ (Variable~\ref{var:ord}),
ordered-triangulation variables $O_{ij}$ (Variable~\ref{var:O}), and width thresholds $W_w$
(Variable~\ref{var:W}).
A width-$w$ query is expressed as
\[
  F(H,w)\ :=\ F_{\text{iter}}(H)\ \wedge\ W_w,
\]
i.e., a SAT call on the fixed CNF $F_{\text{iter}}(H)$ under the single assumption $W_w=1$.
Exact solving is performed either by iterative SAT over $w$ (\textit{iter-lin-up}, \textit{iter-lin-down}, \textit{iter-bin})
or by a single \textit{maxsat} call (Section~\ref{subsec:exact-from-iter}).
We propose two low-overhead refinements that exploit the available heuristics while preserving exactness.

\paperparagraph{Optimization 1: bounds-aware solving.}
\label{opt:bounds-aware-atomic}
Instead of probing widths over the full trivial range, we restrict the backend to the certified
interval $[LB(H),UB(H)]$.
For iterative SAT, this limits the number of tested widths to at most $\Delta+1$ for linear scans and
$\lceil\log_2(\Delta+1)\rceil$ for binary search, where $\Delta = UB(H)-LB(H)$.
For MaxSAT, we keep $F_{\text{iter}}(H)$ as hard clauses and add unit soft clauses only for thresholds
within the interval:
\[
\left(
\underbrace{F_{\text{iter}}(H)}_{\text{\scriptsize hard}},
\ \underbrace{\{\, (W_w,1)\mid w=LB(H),\dots,UB(H)\,\}}_{\text{\scriptsize soft unit clauses, weight }1}
\right).
\]
The empirical impact of bounds-aware solving is reported in
Section~\ref{subsec:exp-sat-enh} (Table~\ref{tab:bounds-aware-speedup}).

\paperparagraph{Optimization 2: warm-starting via phase hints on $\mathit{ord}_{ij}$.}
\label{opt:warmstart-phasehints}
The greedy upper bound also provides an elimination ordering $\prec$, which we translate into phase (polarity) hints for the ordering variables $\mathit{ord}_{ij}$ (Variable~\ref{var:ord})
consistent with $\prec$.
These are solver preferences only: no clauses are added, so satisfiability and optimality are unchanged.
The observed effect is illustrated in Figure~\ref{fig:warmstart-lin-down-myciel4} and discussed in
Section~\ref{subsec:exp-sat-enh}; implementation details are given in Section~\ref{sec:impl-warmstart}.

\subsection{Putting the building blocks together: the HybridSolver recursion}
We obtain the complete solver by combining (i) bounds, (ii) safe-separator decomposition,
(iii) bound propagation, and (iv) an optimized atomic backend that is \emph{bounds-aware}
and supports \emph{warm-starting}. See Algorithm~\ref{alg:hybrid}.

\begin{algorithm}[t]
\caption{\textsc{HybridSolver} recursion (simplified)}
\label{alg:hybrid}
\begin{algorithmic}[1]
\Function{HybridSolve}{$G$}
  \State $(LB,UB,\prec,\mathcal{D}_{UB}) \gets$ \Call{ComputeBounds}{$G$}
  \If{$LB = UB$}
    \State \Return $(UB,\mathcal{D}_{UB})$
  \EndIf
  \State $(S,\{C_1,\dots,C_r\}) \gets$ \Call{FindBestSafeSeparator}{$G$}
  \If{$S \neq \bot$}
    \For{$i=1$ to $r$}
      \State $G_i \gets G[C_i\cup S]\cup K(S)$
      \State $(w_i,\mathcal{D}_i) \gets$ \Call{HybridSolve}{$G_i$}
    \EndFor
    \State $w \gets \max(|S|-1,\max_i w_i)$
    \State $\mathcal{D} \gets$ \Call{MergeDecompositions}{$(\mathcal{D}_i)_i, S$}
    \State \Return $(w,\mathcal{D})$
  \Else
    \State \Return \Call{AtomicSolve}{$G,LB,UB,\prec$}
    \Comment{Berg--J{\"a}rvisalo encoding + bounds-aware + warm-start}
  \EndIf
\EndFunction
\end{algorithmic}
\end{algorithm}

\subsection{Correctness and complexity}
\paperparagraph{Correctness.}
If $LB=UB$, the value is certified by valid lower and upper bounds.
If a safe separator is applied, recursing on the induced subgraphs is correct by
Definition~\ref{def:safe-separator}.
If no separator is found, the atomic backend returns the exact treewidth.
Bounds-aware restriction only excludes widths outside $[LB,UB]$, and warm-starting only provides
solver hints; neither affects satisfiability or optimality. \\

\paperparagraph{Complexity.}
\noindent The per-node overhead is polynomial: bounds computation (MMW + greedy elimination), separator detection
(linear-time routines plus bounded-size enumeration for clique / almost-clique candidates), and
splitting/merging (output-linear in the produced decompositions).
The worst-case runtime remains exponential overall because the atomic SAT/MaxSAT call solves an NP-hard
problem.

\section{Separator-Aware Iterative Encoding}
\label{sec:sep-aware-encoding}

This section introduces a separator-aware iterative encoding that compiles a whole safe-separator
decomposition into a single incremental SAT/MaxSAT instance.
Unlike \textsc{HybridSolver} (Section~\ref{sec:hybrid}), which recurses and invokes an atomic backend at
the leaves, we (i) precompute a safe-separator decomposition tree, (ii) build an independent Berg--J{\"a}rvisalo
iterative encoding for each leaf, and (iii) connect the leaves through a \emph{shared} global width threshold.
Iterative SAT (or MaxSAT) then runs on one global formula, reusing learned clauses across all leaves and all
tested widths.

\subsection{Recursion tree and leaf instances}
We build a safe-separator recursion tree for the input graph $G$ using the same separator
classes as in \textsc{HybridSolver} (Section~\ref{sec:safe-separators}).
Whenever a safe separator $S$ is found in a subgraph $H$, we split $H\setminus S$ into connected
components and recurse on the corresponding separator-augmented subgraphs
$H[C_i\cup S]\cup K(S)$.
This process continues until no further safe separator is found (or a recursion limit is reached),
resulting in leaf graphs $G_1,\dots,G_L$.
By Definition~\ref{def:safe-separator}, cliquifying the separator preserves treewidth, and the
treewidth of the original graph equals the maximum treewidth among the leaf instances.

\subsection{Leaf-local iterative encodings}
For each leaf $\ell\in\{1,\dots,L\}$, we build an independent copy of the Berg--J{\"a}rvisalo iterative
encoding (Encoding~\ref{enc:bj-iter}) on $G_\ell$, and we namespace all variables by $\ell$ so that the
leaf formulas are syntactically disjoint.

\Var{Leaf-local ordering}\label{var:ord-leaf}
For each leaf $\ell$ and $1\le i<j\le n_\ell$, introduce $\mathit{ord}^{\ell}_{ij}$ encoding the local
order $v_i^\ell \prec v_j^\ell$, with derived literals $(\mathit{ord}^{\ell})^*_{ij}$ as in
Variable~\ref{var:ord}.

\Var{Leaf-local ordered-triangulation edges}\label{var:O-leaf}
For each leaf $\ell$ and all distinct $i,j\in\{1,\dots,n_\ell\}$, introduce $O^{\ell}_{ij}$ as in
Variable~\ref{var:O} but restricted to $G_\ell$.

\Var{Leaf-local width-threshold}\label{var:W-leaf}
For each leaf $\ell$ and each $w\in\{0,\dots,n_\ell-2\}$, introduce $W^\ell_w$ with the same incremental
meaning as Variable~\ref{var:W} (width $\le w$ for \emph{that leaf}).

\Enc{Leaf-local iterative encoding $F^\ell_{\text{iter}}(G_\ell)$}\label{enc:bj-iter-leaf}
For each leaf $\ell$, let $F^\ell_{\text{iter}}(G_\ell)$ be Encoding~\ref{enc:bj-iter} instantiated on
$G_\ell$ using Variables~\ref{var:ord-leaf}, \ref{var:O-leaf}, and \ref{var:W-leaf}.
The leaf part of the global CNF is simply
\[
  F_{\text{leaves}}(G,T) \;:=\; \bigwedge_{\ell=1}^{L} F^\ell_{\text{iter}}(G_\ell).
\]

\subsection{Global width variables and leaf linking}
We now introduce one shared width threshold, and force it to agree with the leaves.

\Var{Global width-threshold}\label{var:W-global}
Let $n:=|V(G)|$. Introduce global variables $\widehat{W}_0,\dots,\widehat{W}_{n-2}$, where
$\widehat{W}_w=1$ is intended to mean “all leaves have width $\le w$”. For a fixed $w$, leaves with $n_\ell \le w+1$ are certainly within width $w$ since
$\tw(G_\ell)\le n_\ell-1 \le w$. We therefore only need to constrain the remaining leaves.
Define the active-leaf index set
\[
  \mathcal{L}(w)\ :=\ \{\ell \in \{1,\dots,L\}\mid n_\ell \ge w+2\}.
\]
We enforce the intended semantics by the equivalence
\begin{equation}
\label{eq:sep-aware-linking}
  \widehat{W}_w \;\leftrightarrow\; \bigwedge_{\ell\in \mathcal{L}(w)} W^\ell_w.
\end{equation}

\noindent If $\mathcal{L}(w)=\varnothing$, Equation~\eqref{eq:sep-aware-linking} reduces to the unit clause
$\widehat{W}_w$.

\paperparagraph{Global monotonicity.}
As in the baseline iterative encoding, we add $\widehat{W}_w \rightarrow \widehat{W}_{w+1}$ for all
$w\in\{0,\dots,n-3\}$.

\Enc{Separator-aware iterative CNF $F_{\text{sep-iter}}(G,T)$}\label{enc:sep-aware-iter}
The full separator-aware formula is
\begin{equation*}
\label{eq:sep-aware-full}
\makebox[\linewidth][l]{$\displaystyle
\begin{array}{@{}l@{}}
F_{\text{sep-iter}}(G,T)
\;=\;
\underbrace{F_{\text{leaves}}(G,T)}_{\text{\scriptsize leaf iterative encodings}}
\ \wedge\
\underbrace{\Bigl(\bigwedge_{w=0}^{n-3}(\widehat{W}_w \rightarrow \widehat{W}_{w+1})\Bigr)}_{\text{\scriptsize global monotonicity}}
\\[-0.3mm]
\wedge\
\underbrace{\Bigl(\bigwedge_{w=0}^{n-2}\bigl(\widehat{W}_w \leftrightarrow \bigwedge_{\ell\in \mathcal{L}(w)} W^\ell_w\bigr)\Bigr)}_{\text{\scriptsize link global width to all relevant leaves}}.
\end{array}$}
\end{equation*}

\subsection{Using the encoding for SAT and MaxSAT}
\paperparagraph{Iterative SAT.}
A width-$w$ decision query is a single SAT call on the fixed CNF
$F_{\text{sep-iter}}(G,T)$ under the assumption $\widehat{W}_w=1$.
Monotonicity of $\widehat{W}_w$ supports the same probing strategies as in
Section~\ref{subsec:exact-from-iter} (linear up/down or binary search).  

\paperparagraph{MaxSAT.}
We form a partial MaxSAT instance with all clauses of
$F_{\text{sep-iter}}(G,T)$ hard and unit soft clauses on $\widehat{W}_w$ :
\begin{equation*}
\left(
  \underbrace{F_{\text{sep-iter}}(G,T)}_{\text{\scriptsize hard}},
  \ \underbrace{\{\,(\widehat{W}_w,1)\mid w=0,\dots,n-2\,\}}_{\text{\scriptsize soft}}
\right).
\end{equation*}

\subsection{Solver workflow and decomposition reconstruction}
\label{subsec:sep-aware-solver}

Our separator-aware solver proceeds in two phases: (1) build and solve one global incremental
SAT/MaxSAT instance, and (2) decode a tree-decomposition by combining leaf tree-decompositions along the separators
stored in the recursion tree.

\paperparagraph{Data stored in the recursion tree.}
The safe-separator recursion produces a rooted tree $T$ whose internal nodes store the separator
$S$ used to split the current subgraph and whose edges identify the corresponding separator-augmented
children. Each leaf $\ell$ stores its leaf graph $G_\ell$ and a relabeling map
$\phi_\ell: V(G_\ell^{\text{local}})\rightarrow V(G)$ that translates the leaf’s local vertex ids
(back) to the original vertex names (of the input graph).

\begin{algorithm}[H]
\caption{Separator-aware incremental solving and reconstruction}
\label{alg:sep-aware}
\begin{algorithmic}[1]
\Require Graph $G$
\Ensure Tree-decomposition $\mathcal{D}$ of $G$ of minimum width (via SAT/MaxSAT optimum)
\State Build safe-separator recursion tree $T$ of $G$; obtain leaves $\{G_\ell\}_{\ell=1}^{L}$ and maps $\phi_\ell$
\State Build global CNF $F_{\text{sep-iter}}(G,T)$ (leaf encodings + global $\widehat{W}$ + linking/monotonicity)
\State Find optimum width $w^\star$ by incremental SAT probes (assumptions $\widehat{W}_{w}$) or MaxSAT.
\For{$\ell=1$ to $L$}
  \State $M_\ell \gets$ projection of $M$ onto leaf-$\ell$ variable ids (using the CNF namespacing map)
  \State Extract leaf order $\pi_\ell$ from $\mathit{ord}^\ell$ in $M_\ell$; extract fill edges from $O^\ell$ 
  \State $\mathcal{D}_\ell^{\text{local}} \gets$ build a tree decomposition from $\pi_\ell$ and $O^\ell$ on $G_\ell$ 
  \State $\mathcal{D}_\ell \gets$ map bags of $\mathcal{D}_\ell^{\text{local}}$ through $\phi_\ell$ into $V(G)$
\EndFor
\State \Return $\textsc{MergeUp}(T,\{\mathcal{D}_\ell\})$ \Comment{post-order: add hub bag $S$ at each internal node and connect to children}
\end{algorithmic}
\end{algorithm}

\paperparagraph{Decoding leaf solutions.}
A satisfying assignment of the global CNF fixes, for each leaf $\ell$, an elimination ordering through
the corresponding namespaced ordering variables. For each leaf, we project the model to its local
variables, extract an elimination order, and construct a tree-decomposition
$\mathcal{D}_\ell$, which is then mapped back to the original vertex set.

\noindent The procedure \textsc{MergeUp} traverses the recursion tree bottom-up and combines the leaf
decompositions using the stored separators via \textsc{MergeDecompositions}
(Algorithm~\ref{alg:merge-decompositions}), ultimately producing a tree-decomposition of the original graph.

\subsection{Correctness and complexity}
\label{subsec:sep-aware-correctness}

\paperparagraph{Correctness.}
Let $T$ be the safe-separator recursion tree of $G$ with leaf graphs $G_1,\dots,G_L$, where each split
along a safe separator $S$ recurses on separator-augmented subgraphs $H[C_i\cup S]\cup K(S)$.
By the definition of safe separators, cliquifying $S$ preserves treewidth, and the treewidth of any
subgraph equals the maximum treewidth of its children. Propagating this property through the recursion
yields that $\tw(G)$ is the maximum treewidth among the leaves. \\

\noindent Fix a width bound $w$. For each leaf $\ell$ with at least $w+2$ vertices, the standard iterative
encoding $F^\ell_{\mathrm{iter}}(G_\ell)$ is satisfiable under the assumption $W^\ell_w$ if and only if
$\tw(G_\ell)\le w$. Smaller leaves satisfy the bound trivially and need not be encoded at width $w$.
The linking constraints enforce
\[
\widehat{W}_w \leftrightarrow \bigwedge_{\ell\in\mathcal{L}(w)} W^\ell_w,
\]
ensuring that the global width bound $w$ holds simultaneously for all relevant leaves.
Consequently, the separator-aware encoding is satisfiable under $\widehat{W}_w$ if and only if
$\tw(G)\le w$. Iterating over $w$ therefore identifies the smallest feasible width, which equals $\tw(G)$.
The same reasoning applies to the MaxSAT formulation: monotonicity of the global width variables
ensures that maximizing the number of satisfied width constraints yields the exact treewidth. \\

\paperparagraph{Complexity.}
The size of the separator-aware formula is dominated by the sum of the leaf encodings: we include
one full iterative Berg--J{\"a}rvisalo CNF $F^{\ell}_{\mathrm{iter}}(G_\ell)$ per leaf, and add only
a small global linking layer.
This layer consists of (i) a linear number of monotonicity clauses for the global thresholds
$\widehat{W}_w$, and (ii) linking clauses that connect each $\widehat{W}_w$ to the active leaves
at width $w$.
In the worst case, this contributes $O(nL)$ additional clauses, where $n=|V(G)|$ and $L$ is the
number of leaves.  \\

\noindent The contribution of the flattened separator recursion overhead is polynomial, while the overall worst-case runtime of course remains exponential, as exact treewidth is NP-hard; the goal of
the construction is to let one incremental SAT/MaxSAT
run share learned information across all leaf subinstances and across all tested widths. Note also that on the reconstruction side, all steps run in polynomial time: each leaf is decoded into a
tree-decomposition in polynomial time (Appendix~\ref{sec:impl-decode-td}), and the resulting leaf
decompositions are merged bottom-up along the recursion tree in polynomial time (Algorithm~\ref{alg:merge-decompositions}).

\clearpage
\section{Experimental Evaluation}
\label{sec:experiments}

This section evaluates the solvers introduced in
Section~\ref{sec:baseline-sat} (\textsc{Regular}),
Section~\ref{sec:hybrid} (\textsc{HybridSolver}), and
Section~\ref{sec:sep-aware-encoding} (\textsc{Separator-aware})
across several benchmark settings, focusing on runtime, solved instances under time limits, and
the practical impact of exploiting safe-separator structure. All experiments were run on a desktop machine with an Intel Core i7-10700K (8C/16T) and 64\,GB RAM (CachyOS / Arch Linux). \\
\noindent\textbf{General Experimental Setting.} \label{general:experimental_setting} Unless stated otherwise, we compare the three configurations \emph{Regular} which is our baseline, \emph{Hybrid}, and
\emph{Separator-aware} throughout. Each configuration is evaluated under four exact modes:
\texttt{maxsat}, \texttt{iter-bin}, \texttt{iter-lin-up} and \texttt{iter-lin-down}. Unless stated otherwise, results are reported per mode and/or aggregated across modes. All SAT-based runs use \textsc{Glucose}~4 as the underlying SAT solver ~\cite{audemard2009predicting,glucose-web},
and MaxSAT runs use \textsc{RC2} as the optimizer backend (internally calling a SAT solver).
Our implementation also supports \textsc{MiniSat}~2.2 ~\cite{een2003extensible} and can be extended with minimal effort to other
incremental SAT solvers, as solver interaction is isolated behind a thin backend wrapper. We report results on four settings:
(1) an instance set closely matching the original Berg--J{\"a}rvisalo benchmark, with some differences due to unavailable graphs and version discrepancies across sources;
(2) pseudo-random clustered graphs;
(3) Erd\H{o}s--R\'enyi random graphs; and
(4) the PACE2017 exact treewidth benchmark. For randomized synthetic benchmarks, we keep track of the random seeds for reproducibility.
In addition to end-to-end comparisons, we also evaluate in isolation (i) the two SAT-layer enhancements used by the HybridSolver for
atomic backend ---\emph{bounds-aware solving} and \emph{warm-start hinting}---and
(ii) the usefulness of the polynomial-time heuristic bounds as standalone guidance/pruning signals.

\subsection{Berg--J\"arvisalo Benchmark}
\label{sec:exp-berg}

We first evaluate on the benchmark used by Berg and J\"arvisalo to assess SAT-based treewidth
computation~\cite{BergJarvisalo2014SATTreewidth}.
Following their setup, we consider a 34-graph collection gathered from standard treewidth instance
archives and Bayesian--network repositories, including \texttt{libtw.zip} ~\cite{twlib-libtw-zip},
\texttt{coloring.zip} ~\cite{twlib-coloring-zip}, and the \texttt{bnlearn} repository ~\cite{bnlearn-bnrepository}.
These instances cover a range of graph structures at moderate sizes and are a widely used reference
point for comparing SAT/MaxSAT-based exact treewidth solvers.\\

\noindent \textbf{Experimental Setting.}
For each instance graph, we run each configuration on all modes (following the general experimental setting ~\ref{general:experimental_setting}), 
with a timeout of 10 minutes per run. Runs exceeding the limit are counted as timeouts.\\

\begin{figure}[t]
  \centering
  \includegraphics[width=1.0\linewidth]{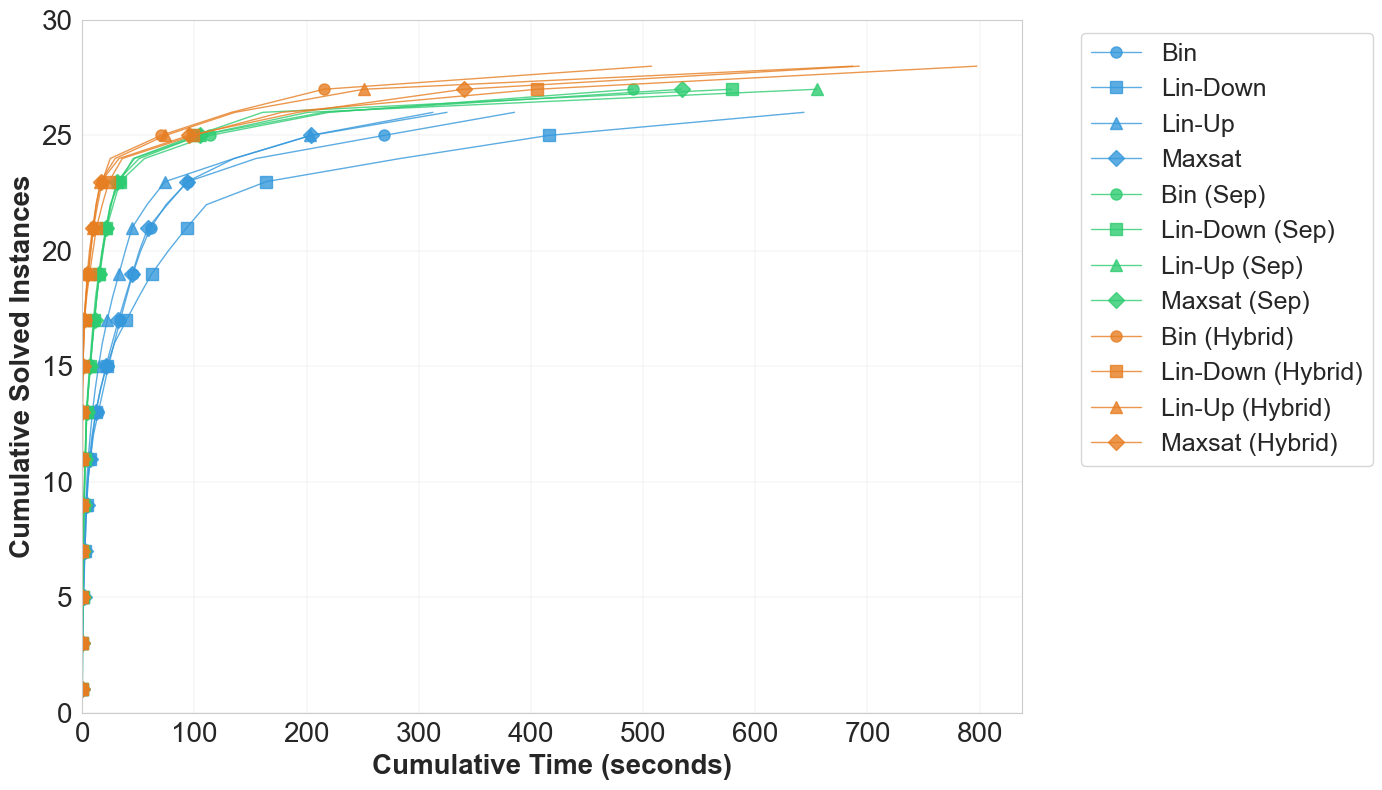}
  \caption{Cumulative number of solved Berg--J\"arvisalo benchmark instances over time.}
  \label{fig:berg-cactus}
\end{figure}

\noindent \textbf{Results.}
Figure~\ref{fig:berg-cactus} plots the cumulative number of solved instances over time.
Both separator-based approaches dominate the Regular baseline and solve more instances without timing out, and this, independently of which underlying mode is used. 
\emph{Hybrid} consistently reaches higher solved counts earlier and remains the most robust near the
timeout, while \emph{Separator-aware} also improves substantially over \emph{Regular}, confirming that
exposing safe-separator structure yields practical gains even within a SAT/MaxSAT pipeline. Full raw results, pairwise comparison and aggregated results are provided in Appendix~\ref{app:berg-extended}.

\subsection{Pseudo-Random Clustered Graphs}
\label{sec:exp-clustered}

We next evaluate on a synthetic benchmark designed to exhibit strong separator structure. 
Instances consist of $k$ dense clusters of size $n$ connected by a sparse inter-cluster backbone,
so that articulation points and small clique-like separators occur frequently (see Appendix~\ref{subsec:app-clustered-example} for more details and representative example instance). 
Such clustered graphs model systems composed of tightly connected components with sparse
interconnections, a structure common in many real-world networks and graphical models.  

\noindent \textbf{Experimental Setting.}
We generate instances over a grid of parameters $(k,n)$ using fixed random seeds, where k (numbers of clusters) ranges from 2 to 14 and n (cluster sizes in amounts of vertices) from 5 to 20. Each run is repeated five times per fixed $(k,n)$, and we report mean wall-clock runtime; runs exceeding the timeout of 60 seconds are counted as timeouts. Note that a grid is considered as timed-out if a majority out of the 5 instances times out for given configuration and associated mode. 

\begin{figure*}[t]
  \centering
  \includegraphics[width=1.0\textwidth]{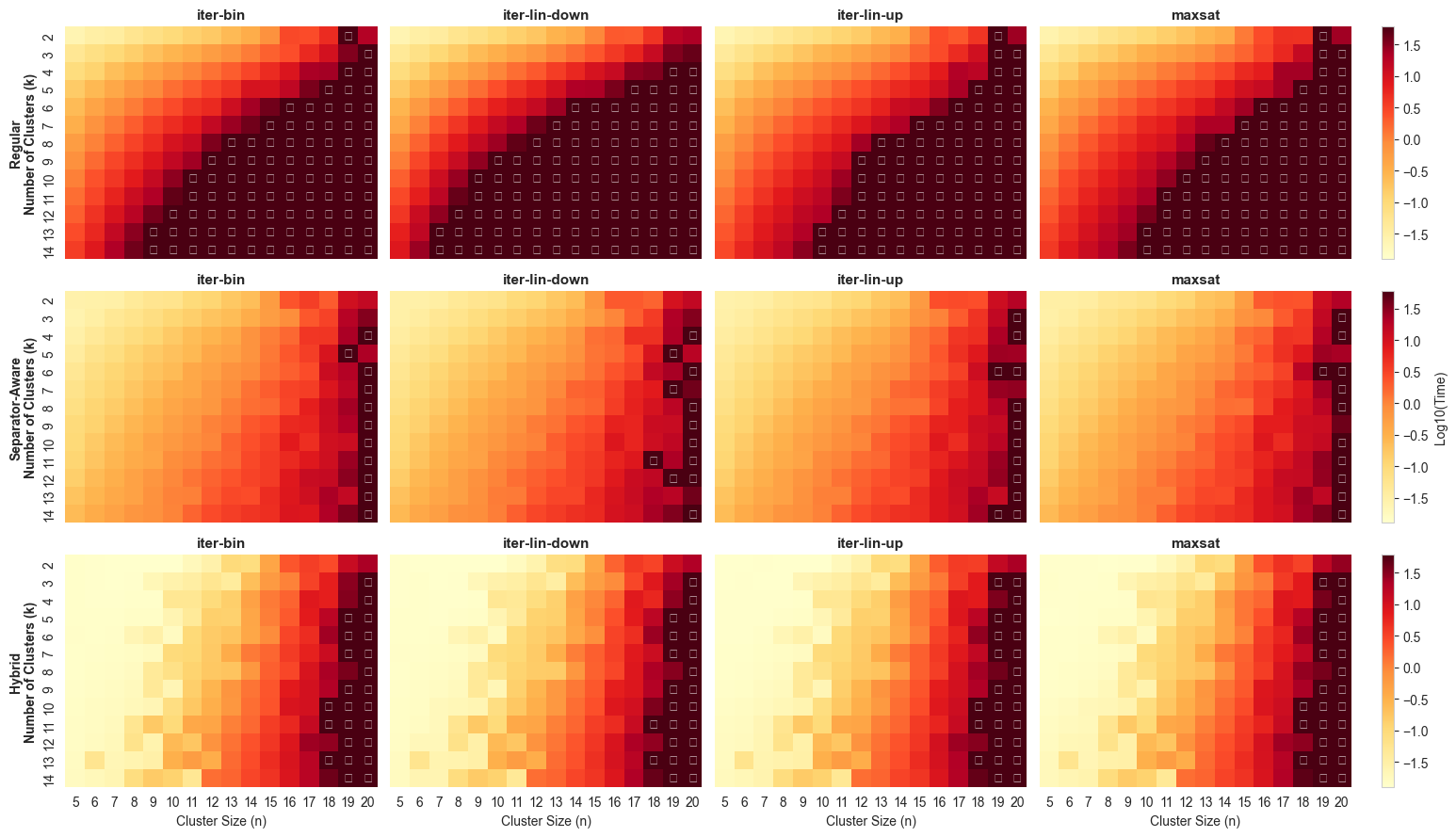}
  \caption{Pseudo-random clustered benchmark: runtime heatmaps (log$_{10}$ seconds) over cluster size $n$
  and number of clusters $k$. White markers denote timeouts.}
  \label{fig:clustered-heatmap}
\end{figure*}

\noindent \textbf{Results.}
Figure~\ref{fig:clustered-heatmap} shows runtime heatmaps (log$_{10}$ seconds) across the $(k,n)$ grid.
Both separator-based approaches substantially outperform the \emph{Regular} baseline, confirming that safe
separators capture the dominant structure in these clustered instances and shrink the hardest SAT/MaxSAT
subproblems. The improvement is often by orders of magnitude: large regions that time out for
\emph{Regular} become solvable in seconds for \emph{Hybrid} and \emph{Separator-aware}. Interestingly, \emph{Separator-aware} can even outperform \emph{Hybrid} on the hard right-hand side of the grid:
there are parameter regions where \emph{Hybrid} still hits the time limit while
\emph{Separator-aware} completes it. A plausible explanation is that the separator-aware encoding exposes more reuse to the SAT/MaxSAT solver.
\textsc{HybridSolver} prunes primarily at the algorithmic level by decomposing the instance and tightening bounds,
but the SAT work performed in each leaf remains largely isolated across recursive calls.
In contrast, the separator-aware approach compiles all leaves into a single \emph{incremental} SAT/MaxSAT instance
with a shared global width threshold, so learned clauses and propagations can be reused across leaves and across repeated width probes.
As a result, a global width-$w$ query can be refuted as soon as one leaf forces width $>w$, without fully exploring the remaining leaves,
which explains why separator-aware can outperform \emph{Hybrid} on the hardest region of the grid.

\subsection{Random Graphs}
\label{subsec:exp-random}

We next evaluate on Erd\H{o}s--R\'enyi ~\cite{ErdosRenyi1959} random graphs $G(n,p)$, which let us vary density in a
controlled way while largely avoiding the explicit community structure of the clustered benchmark.
As $p$ increases, these graphs transition from sparse (often disconnected) instances with many
trivial separator opportunities to well-connected instances where separator structure becomes rare
and treewidth grows quickly. \\

\noindent
\textbf{Experimental Setting.}
We generate a grid of Erd\H{o}s--R\'enyi instances $G(n,p)$ with fixed seeds, using
$n$ in the range $30$--$80$ vertices and edge probabilities $p$ in the range $1\%$--$4\%$.
For each parameter setting we sample five graphs and report average wall-clock runtime.
Each instance is solved under all configurations and all modes. We record runtime
and count runs exceeding the per-run timeout of 60 seconds as timeouts. \\

\begin{figure}[H]
  \centering
  \includegraphics[width=0.9\linewidth]{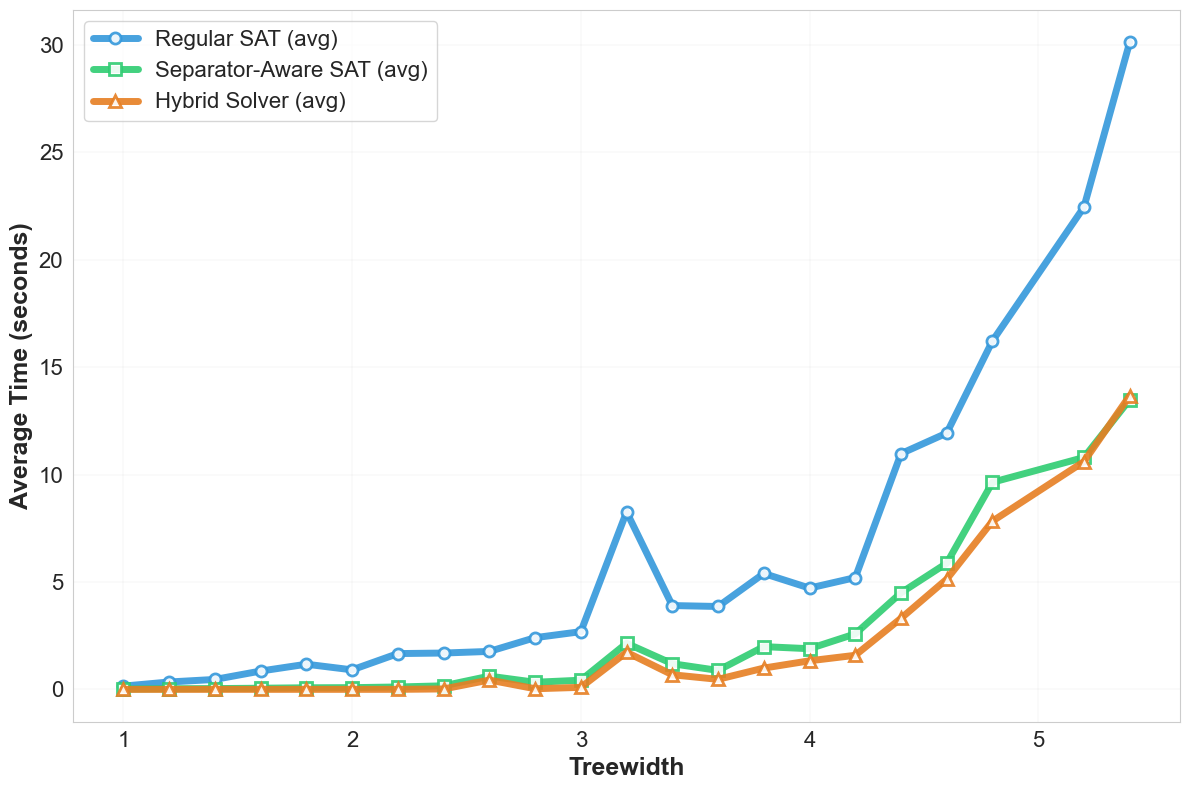}
  \caption{Average runtime versus mean treewidth per $(n,p)$ setting on Erd\H{o}s--R\'enyi instances, averaged across modes.}
  \label{fig:random-treewidth}
\end{figure}

\noindent
\textbf{Results.}
To summarize behavior beyond raw $(n,p)$ parameters, we group parameter settings by the mean optimal
treewidth of their five graphs and average runtimes across modes (keeping only settings where all runs succeeded). Figure~\ref{fig:random-treewidth} reports average runtime against this mean treewidth.
All three configurations exhibit the same overall dependence on treewidth in this setting: runtimes
remain near-zero for very small widths and then increase steeply once the treewidth reaches the
moderate range, reflecting that treewidth itself is the dominant driver of difficulty for these
Erd\H{o}s--R\'enyi instances. Within this shared trend, the separator-based variants consistently improve upon the
\emph{Regular} encoding, indicating that structure-aware
reasoning reduces solving time even on random graphs. It is worth noting that when treewidth increases, all solvers' runtime increases (follows similar trend), suggesting that once the instances become intrinsically hard (high
width), all configurations are increasingly constrained by the same underlying SAT/MaxSAT effort.

\noindent See Appendix for full grid results detailed in ~\ref{app:random-extended}.

\subsection{SAT-level enhancements}
\label{subsec:exp-sat-enh}

We isolate the two SAT-layer optimizations (Section \ref{subsec:sat-enhancement-building-block}) used by \textsc{HybridSolver}
when it can no longer split a graph and has to rely on the underlying backend (which is the
\emph{Regular} Berg--J{\"a}rvisalo routine). \\

\noindent We therefore evaluate the effect of: 
(i) \emph{bounds-awareness}, which restricts the explored widths to a certified interval $[LB,UB]$ rather than
$[1,|V|{-}1]$, and (ii) \emph{warm-starting}, which injects polarity preferences for the ordering variables
$\mathit{ord}_{ij}$ (Variable~\ref{var:ord}) derived from a heuristic elimination ordering, without adding
constraints.
 
\medskip\noindent\textbf{Bounds-aware solving.}
On the \texttt{david} graph from the Berg--J\"arvisalo benchmark, we evaluate bounds-aware solving by
running each exact mode five times with and without initial bounds.
This instance has $n=87$ vertices, so without bounds the atomic search ranges over widths
$w\in\{1,\dots,|V|{-}2\}=\{1,\dots,85\}$.
With heuristic bounds enabled, the initial window is $[LB,UB]=[10,13]$ and the true treewidth is $\tw=13$.
Table~\ref{tab:bounds-aware-speedup} reports mean wall-clock runtimes and resulting speedups,
showing consistent improvements across all four modes, with the largest gains for
\texttt{iter-lin-down} and \texttt{maxsat}.

\begin{table}[H]
\caption{Effect of initial bounds on \texttt{david}: mean runtime (s) over five runs and speedup
($LB=10$, $UB=13$, $\tw=13$).}
\label{tab:bounds-aware-speedup}
\centering
\begin{tabular}{lrrr}
\toprule
\textbf{Strategy} & \textbf{Without bounds} & \textbf{With bounds} & \textbf{Speedup} \\
\midrule
\texttt{iter-bin}   & 58.797 & 38.702 & 1.52$\times$ \\
\texttt{iter-lin-u} & 59.843 & 54.701 & 1.09$\times$ \\
\texttt{iter-lin-d} & 76.322 & 39.200 & 1.95$\times$ \\
\texttt{maxsat}     & 69.464 & 41.102 & 1.69$\times$ \\
\bottomrule
\end{tabular}
\end{table}

\begin{figure}[t]
  \centering
  \includegraphics[width=\linewidth]{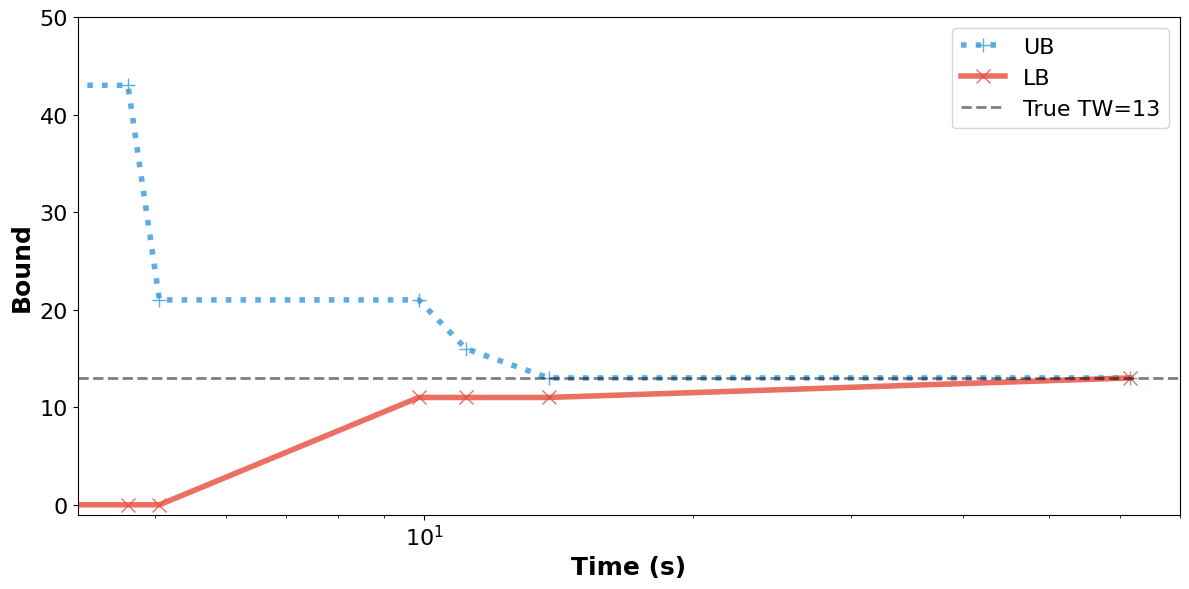}
  \caption{Binary search bounds over time for \texttt{david.gr} without initial bounds.}
  \label{fig:binsearch-bounds-over-time}
\end{figure}

\noindent Nevertheless, Figure~\ref{fig:binsearch-bounds-over-time} illustrates the typical runtime profile of binary
search over the width: while the interval shrinks quickly at first, the expensive iterations concentrate near the
optimum, where further narrowing becomes progressively harder.
Bounds-aware solving therefore primarily removes avoidable work away from the optimum (as usually the heuristic bounds are quite tight), but it does
not eliminate the intrinsic difficulty of proving (or refuting) widths close to the true treewidth,
which is often the dominant cost in exact treewidth computation.
This behavior matches the trends reported by Berg and J\"arvisalo ~\cite{BergJarvisalo2014SATTreewidth} for iterative
SAT-based treewidth solving, where runtime is likewise dominated by the final iterations near the
true width.

\medskip \noindent\textbf{Warm-start hinting.}
Warm-starting provides the SAT solver with an initial \emph{hint} biased toward a promising
elimination ordering---in our case derived from the greedy upper-bound ordering---with the goal of
steering early branching decisions in a favorable direction. In practice, its impact is highly
instance-dependent: it can yield slight speedups with high variance, but often has little to no
effect. A key reason is that the hint is not a hard constraint: modern SAT solvers (e.g.,
\textsc{Glucose}) treat phase suggestions as preferences and may ignore or override them as search
progresses and conflicts accumulate. We provide an illustrative example on \texttt{myciel4.gr} in
Figure~\ref{fig:warmstart-lin-down-myciel4} (Appendix), where \texttt{iter-lin-down} improves
slightly; overall, warm-starting should be viewed as a low-cost, opportunistic optimization rather than a
mechanism that drastically reduces overall runtime.

\subsection{Bound heuristics accuracy}
\label{sec:exp-bounds}

As described in Section \ref{subsec:hybrid-bounds},  \textsc{HybridSolver} uses upper and lower bounds on treewidth to both prune and guide exact solving. 
We therefore evaluate the quality of the two bound heuristics used throughout \textsc{HybridSolver} in isolation.

\medskip\noindent
\textbf{Experimental Setting.}
We evaluate bounds on 200 instances from the PACE2017 exact treewidth benchmark
(\emph{regular} and \emph{bonus} instances), for which optimal treewidth values are available.
Instance sizes range from 48 to 3{,}706 vertices (mean 383.9) and span diverse structures, where associated treewidth ranges from 6 to 908 with a mean value of 34.48.

\begin{figure*}[t]
  \centering
  \includegraphics[width=1.0\textwidth]{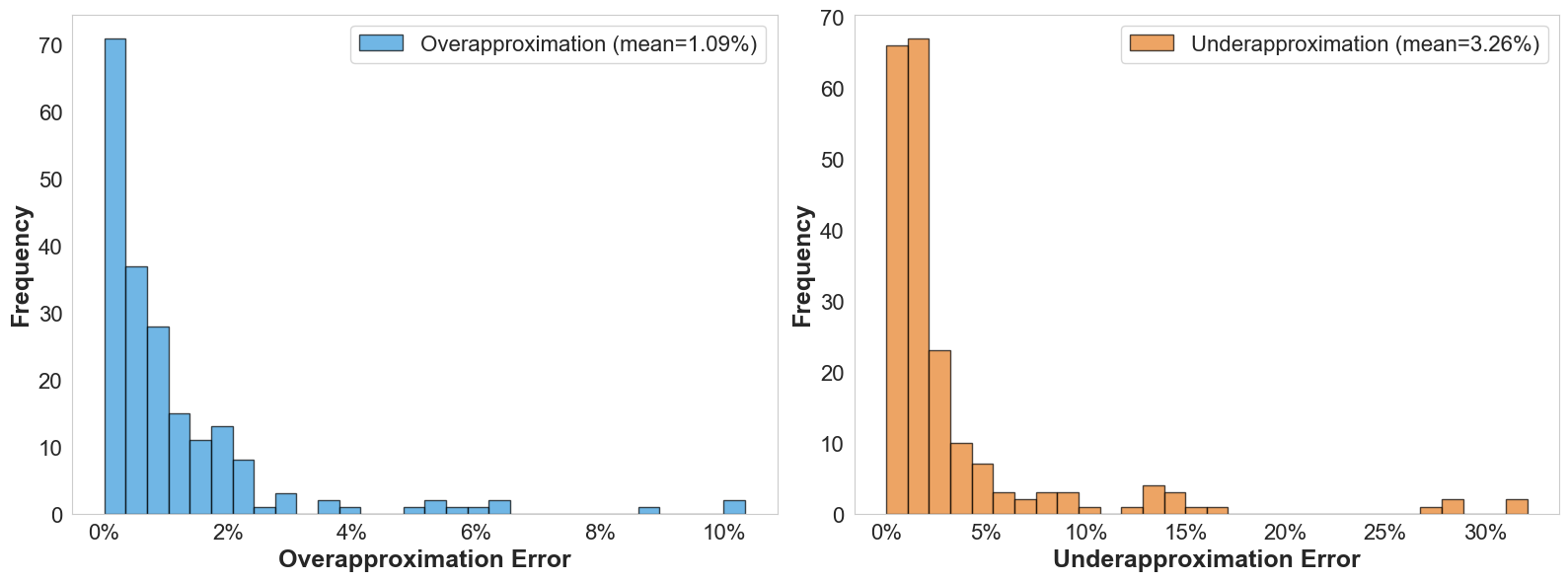}
  \caption{Normalized error distributions of bound heuristics on PACE2017 instances.
           Left: upper-bound error $e_{\mathrm{UB}}(G)/n$.
           Right: lower-bound error $e_{\mathrm{LB}}(G)/n$.}
  \label{fig:bounds-errors}
\end{figure*}

\medskip\noindent
\textbf{Results.}
Figure~\ref{fig:bounds-errors} reports the distribution of normalized errors
$e_{\mathrm{UB}}(G)/n$ and $e_{\mathrm{LB}}(G)/n$ over the PACE2017 instances.
The greedy upper bound is typically close to optimal (most mass near zero), while the lower
bound is looser and exhibits a longer tail.
Nevertheless, both bounds remain reasonably tight relative to graph size on the majority of
instances, and in a nontrivial fraction of cases they coincide, allowing \textsc{HybridSolver} to
solve the instance without having to invoke the SAT/MaxSAT backend.

\medskip\noindent

\noindent Overall, these results confirm that simple polynomial-time bounds provide strong guidance in
practice, consistent with prior work on treewidth heuristics \cite{BodlaenderKoster2010UpperBounds,BodlaenderKoster2011LowerBounds}.
Inside \textsc{HybridSolver}, even when bounds do not meet exactly, they substantially narrow the
hard width region that the atomic SAT/MaxSAT backend must explore, improving pruning, early
termination, and focus on the intrinsically difficult regime near the true treewidth. See Appendix~\ref{app:bounds-extended} for more details.

\subsection{PACE2017}
\label{sec:exp-pace}

We finally consider the PACE2017 exact treewidth benchmark, which serves as a test for highly specialized competitive exact treewidth solvers. We focus on the regular
instances (odd-numbered: \texttt{ex001}, \texttt{ex003}, \dots), yielding $100$ graphs with
treewidths ranging from $6$ to $908$ (mean $37.8$), and vertex counts
$\min |V|=48$, $\max |V|=3706$, with an average of $\approx 413$ vertices.

\medskip\noindent
\textbf{Experimental setup.}
We evaluate the three SAT-based configurations introduced earlier—\emph{Regular},
\emph{Separator-aware}, and \emph{HybridSolver}—but this time under the same mode \emph{iter-bin} (iterative binary search over the width). Each graph instance is run once per configuration with a
timeout of 5 minutes. For context, we also report results for the two top-performing PACE2017 solvers:
Tamaki's positive-instance driven solver (PACE2017 1st place)~\cite{Tamaki2017PositiveInstanceDriven}
and \textsc{Jdrasil} (PACE2017 2nd place)~\cite{bannach_et_al:LIPIcs.SEA.2017.28}, rerun under the
same hardware and timeout policy.

\medskip\noindent
\textbf{Results.}
Among the SAT-based configurations, only \textsc{HybridSolver} solves any PACE2017 instance to optimality within the time limit, while the PACE2017 winning solvers solve most instances without timing out. This is expected given the goal of the benchmark: PACE2017 is designed to separate treewidth-specific engines that rely on highly domain-guided strategies from SAT-based approaches that encode the problem and delegate solving to a SAT solver. Notably, the few instances solved by \textsc{HybridSolver} are precisely those where it can prune instantly because the heuristic bounds coincide ($LB=UB$), certifying optimality without invoking the SAT/MaxSAT backend. Beyond these cases, once the solver must fall back to the atomic SAT backend, the remaining subproblems are typically too large, and runs time out. In most PACE2017 instances, safe-separator decomposition yields only limited and often unbalanced reductions, leaving a dominant residual leaf that remains beyond the reach of the atomic backend.\\

\begin{figure}[t]
  \centering
  \includegraphics[width=1.0\linewidth]{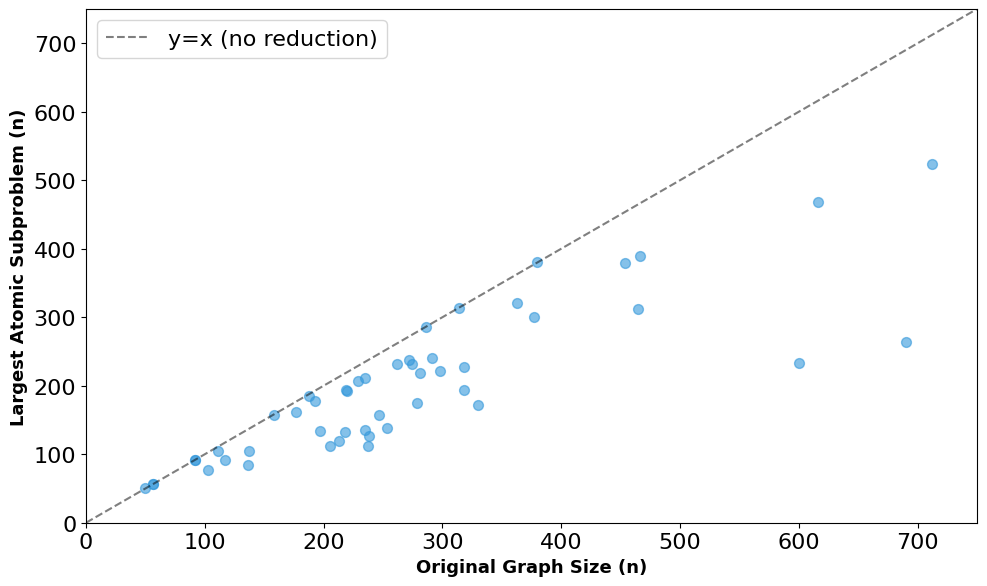}
  \caption{PACE2017: size of the largest atomic leaf subproblem after safe-separator decomposition
  versus the original graph size. The dashed line $y=x$ indicates no reduction.}
  \label{fig:pace-largest-atomic}
\end{figure}

\noindent To quantify how much safe-separator decomposition reduces the instance before having to rely on our SAT/MaxSAT backend on those specific graphs from PACE2017,
Figure~\ref{fig:pace-largest-atomic} plots the size of the largest remaining atomic leaf 
subproblem (graph which cannot be decomposed further) versus the original graph size (points below $y=x$ indicate reduction). While many
instances admit nontrivial reductions, these are often unbalanced, leaving one dominant residual
subinstance close to the original size—hence the limited impact on overall solvability at PACE scale.

\section{Discussion and Limitations}
\noindent SAT- and MaxSAT-based exact treewidth solving can benefit substantially by incorporating graph structure, but the
empirical experiments also clarify where the approach remains constrained. When instances admit meaningful safe-separator
decompositions, reducing the size of the hardest residual subproblem can yield orders-of-magnitude speedups, as most clearly
seen on clustered graphs where both \emph{Hybrid} and \emph{Separator-aware} dominate the \emph{Regular} Berg--J{\"a}rvisalo
solver. 

\medskip
\noindent Comparing the two integrations highlights a trade-off. \textsc{HybridSolver} uses decomposition as control flow and
combines it with bounds and bounds-aware atomic solving, whereas the separator-aware approach compiles the full
decomposition into one global incremental SAT/MaxSAT instance. With many leaves and repeated width checks, this global
coupling can benefit from clause reuse and solver-internal pruning across leaves, and a global width-$w$ query can fail as
soon as one leaf forces width $>w$, which helps explain why separator-aware can outperform \textsc{HybridSolver} on the
hardest clustered instances.

\medskip
\noindent The main limitation is that when separator structure is scarce or highly unbalanced, both decomposition-based
variants increasingly face the same bottleneck as the baseline encoding. This occurs on denser random graphs and on the
PACE2017 instances, where safe separators often remove small fragments while leaving one dominant residual subgraph close to
the original size, thus being equivalent to having to rely on the original Berg--J{\"a}rvisalo encoding, for which those instances are beyond feasibility. Consequently, while safe separators are highly effective on structured
instances, the method remains limited when graphs do not decompose meaningfully.

\section{Future Work}
\noindent Beyond the improvements reported here, several other alternatives were explored but were not yet
conclusive, and therefore remain open directions.

\medskip
\noindent First, we ran early trials to exploit \emph{parallelism} in the iterative SAT workflow by solving multiple
width queries concurrently (e.g., testing several values of $w$ in parallel).
This was not consistently beneficial: parallel execution requires separate solver instances, which forfeits
incremental solving and learned-clause reuse across successive width probes, often increasing total effort.

\medskip
\noindent Second, we experimented with exploiting a good heuristic elimination ordering more aggressively than warm-start phase hints. Concretely, the idea was to fix the first $k$
positions of the ordering to match a heuristic prefix by directly constraining the ordering variables $\mathit{ord}_{ij}$
(i.e., enforcing $\mathit{ord}_{uv}=1$ for pairs $(u,v)$ whose relative order is implied by the fixed prefix), and then let
the SAT solver complete the remaining suffix, effectively searching over all completions consistent with a promising start in
the hope of improving the upper bound.
In preliminary trials this did not yield consistent gains: fixing such $\mathit{ord}_{ij}$ variables can easily overconstrain
the formula and push the solver into harder regions of the search space, and a prefix that is locally good for the heuristic
is not necessarily compatible with an optimal global ordering. Still, this direction seems worth revisiting if one can
identify conditions under which committing to a greedy prefix is provably safe.

\medskip
\noindent Third, we attempted to exploit separators of unknown safeness to force additional decomposition when the graph is
effectively unbreakable with respect to the safe-separator classes used in this work. For instance, we tried selecting small
vertex cuts (e.g., min-cut separators) and cliquifying them, treating them as if they were safe. This approach was not
effective in practice and comes with two major pitfalls. The first is algorithmic: committing to an unsafe separator is hard
to undo, and exploring multiple candidates with backtracking is not viable here because each downstream SAT/MaxSAT call can be
expensive even on reduced instances, making branching over separator choices unfeasible in practice. The second is quality: cliquifying a
separator that is not safe typically induces a substantial increase in width, often producing a loose over-approximation that
is not a useful upper bound. Nevertheless, it may be possible to make progress on this direction by designing verifiable safety
tests or by integrating such cuts only as optional
guidance rather than as committing transformations, which remains an open direction for future work.

\section{Conclusion}
\noindent This report studied SAT- and MaxSAT-based exact treewidth computation with an emphasis on exploiting graph
structure. Building on the Berg--J{\"a}rvisalo encoding as an atomic backend, it integrated safe separators in two ways:
a recursive meta-solver (\textsc{HybridSolver}) combining bounds, decomposition, warm-start and bounds-aware solving, and a
separator-aware iterative encoding that compiles a full decomposition into one incremental SAT/MaxSAT instance by
namespacing leaf encodings and linking their width thresholds globally.

\medskip
\noindent Experiments confirm that the Berg--J{\"a}rvisalo-style SAT/MaxSAT backend is a strong and competitive starting
point: it provides a robust atomic solver on moderate instances and, crucially, it benefits noticeably from the proposed
extensions. Across all evaluated settings, including clustered and Erd\H{o}s--R\'enyi random graphs, explicitly leveraging
graph structure within the SAT/MaxSAT framework leads to substantial gains whenever the instance decomposes
well, frequently turning baseline timeouts into fast solves. The separator-aware variant can further benefit from solver reuse
and cross-leaf pruning when many leaves and repeated width checks occur.

\medskip
\noindent Overall, the results highlight a useful interplay between structure and generality in SAT-driven treewidth. SAT/MaxSAT encodings provide a uniform, solver-agnostic core that is easy to reuse and extend, but they benefit strongly from making graph structure explicit at the instance level.

\bibliographystyle{ACM-Reference-Format}
\bibliography{references}

\clearpage
\appendix
\section*{Appendix}

\addcontentsline{toc}{section}{Appendix}

\section{Extended Experimental Evaluation: Berg--J\"arvisalo Benchmark}
\label{app:berg-extended}

This appendix provides additional results and information for the the Berg--J\"arvisalo benchmark beyond what is presented in the main paper (Section~\ref{sec:exp-berg}).

\subsection{Aggregate performance metrics}
\label{subsec:exp-berg-aggregate}

\begin{figure}[H]
  \centering
  \includegraphics[width=\linewidth]{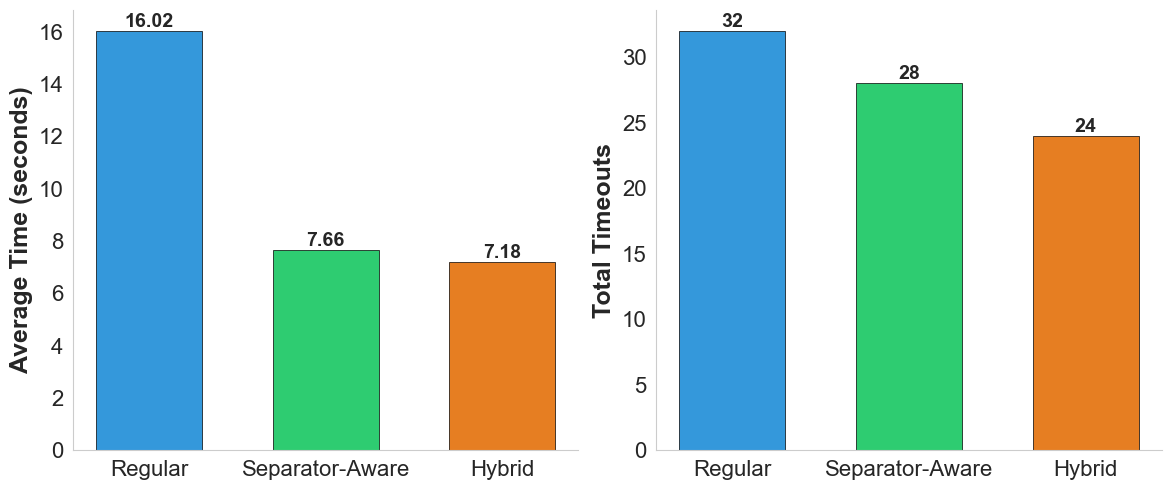}
  \caption{Aggregate performance metrics on the Berg--J\"arvisalo benchmark.
           Left: average time on successful instances.
           Right: number of timeouts.}
  \label{fig:berg-metrics}
\end{figure}

Figure~\ref{fig:berg-metrics} summarizes aggregate performance statistics: average solving time over solved instances (by all configurations and modes), and the total number of timeouts.

\noindent Overall, \textsc{Hybrid} performs best: it solves the largest number of instances (fewest timeouts)
and is also the fastest on successful runs. For the runtime comparison, we report the average over
the subset of instances solved by all configurations, so times are directly comparable.
\emph{Separator-aware} comes second and already yields a clear improvement over the \emph{Regular}
baseline, confirming that exploiting safe-separator structure substantially helps in practice.
Finally, \emph{Hybrid} retains an additional advantage thanks to the early pruning capacities stemming from the upper-bound being equal to the lower-bound.

\subsection{Pairwise solver comparison}
\label{subsec:exp-berg-pairwise}

\begin{figure}[t]
  \centering
  \includegraphics[width=0.9\linewidth]{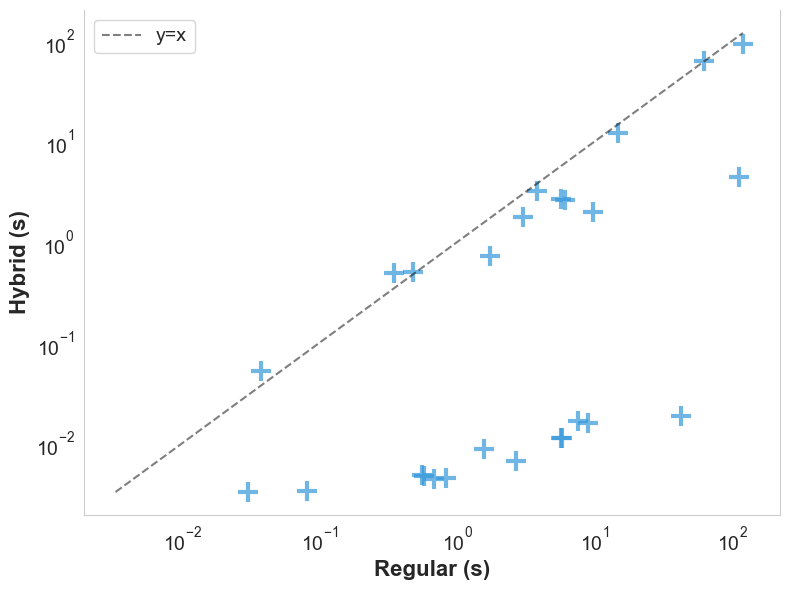}
  \caption{\emph{Hybrid} vs.\ \emph{Regular}.}
  \label{fig:berg-pairwise-hybrid-regular}
\end{figure}

\begin{figure}[t]
  \centering
  \includegraphics[width=0.9\linewidth]{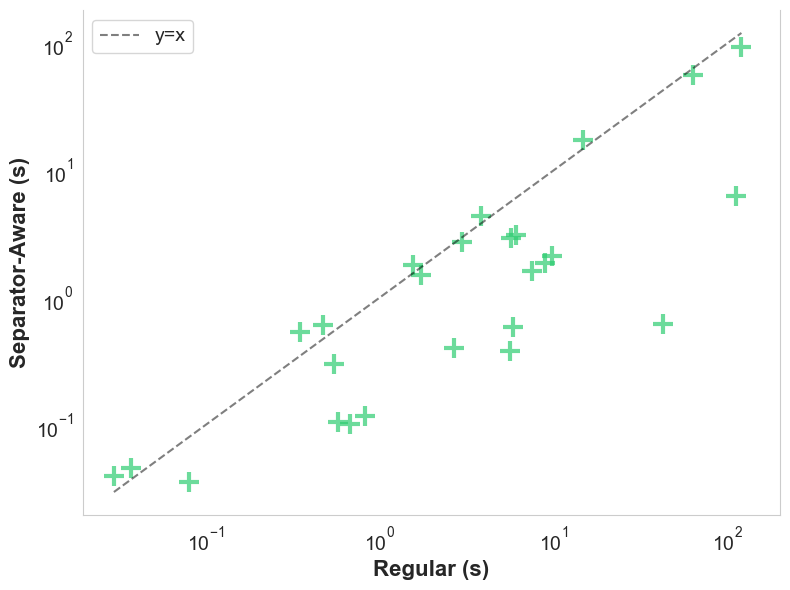}
  \caption{\emph{Separator-aware} vs.\ \emph{Regular}.}
  \label{fig:berg-pairwise-sep-regular}
\end{figure}

\begin{figure}[t]
  \centering
  \includegraphics[width=0.9\linewidth]{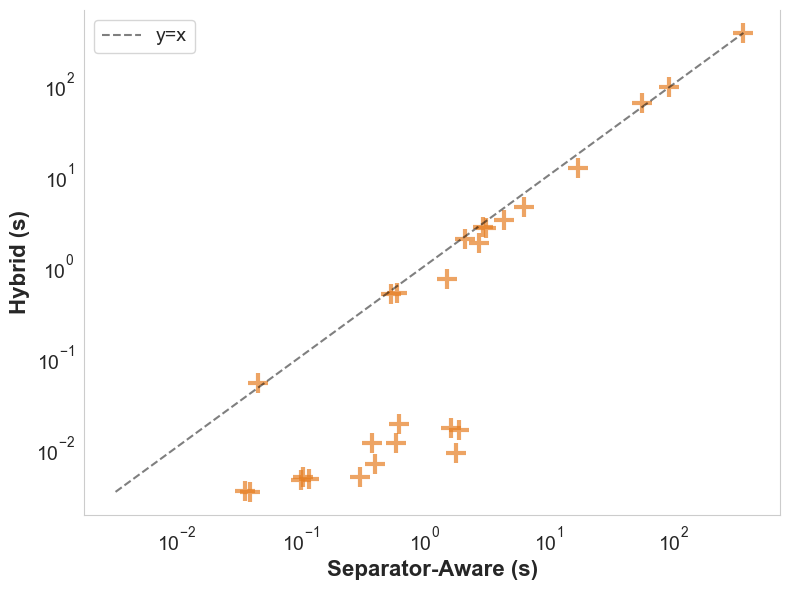}
  \caption{\emph{Hybrid} vs.\ \emph{Separator-aware}.}
  \label{fig:berg-pairwise-hybrid-sep}
\end{figure}

Figures~\ref{fig:berg-pairwise-hybrid-regular}--\ref{fig:berg-pairwise-hybrid-sep} compare solving
times instance-wise on a log--log scale (diagonal $y=x$). For each instance and configuration, the
plotted time is the \emph{average} over the four solving modes
(\texttt{iter-bin}, \texttt{iter-lin-up}, \texttt{iter-lin-down}, \texttt{maxsat}), considering only
successful runs. Points below the diagonal favor the \emph{y}-axis configuration. \\

\noindent Both \textsc{Hybrid} and \textsc{Separator-aware} outperforms \textsc{Regular}
(Figures~\ref{fig:berg-pairwise-hybrid-regular} and~\ref{fig:berg-pairwise-sep-regular}). In
Figure~\ref{fig:berg-pairwise-hybrid-sep}, points concentrate in the lower-left (instances solved
quickly by both), yet \textsc{Hybrid} is slightly faster more often, which we attribute to earlier
pruning enabled by heuristic lower/upper bounds even when separator decomposition already helps.

\begin{table*}[t]
\centering
\scriptsize
\setlength{\tabcolsep}{3pt}
\resizebox{\textwidth}{!}{%
\begin{tabular}{l r r r | r r r r | r r r r | r r r r}
\toprule
\multirow{2}{*}{Instance} & \multirow{2}{*}{TW} & \multirow{2}{*}{$|V|$} & \multirow{2}{*}{$|E|$} & \multicolumn{4}{c|}{Regular} & \multicolumn{4}{c|}{Separator-Aware} & \multicolumn{4}{c}{Hybrid} \\
 &  &  &  & Iter-Lin-U & Iter-Lin-D & Iter-Bin & MaxSAT & Iter-Lin-U & Iter-Lin-D & Iter-Bin & MaxSAT & Iter-Lin-U & Iter-Lin-D & Iter-Bin & MaxSAT \\
\midrule
alarm & 4 & 37 & 65 & 0.45 & 0.67 & 0.53 & 0.64 & $\mathbf{<0.1}$ & 0.11 & 0.11 & 0.11 & $\mathbf{<0.1}$ & $\mathbf{<0.1}$ & $\mathbf{<0.1}$ & $\mathbf{<0.1}$ \\
anna-pp & 12 & 22 & 148 & 2.51 & \textbf{0.93} & 5.06 & 3.46 & 4.30 & 1.81 & 2.79 & 1.98 & 1.35 & 3.07 & 1.35 & 1.33 \\
barley & 7 & 48 & 126 & 5.64 & 5.81 & 5.56 & 5.63 & 2.95 & 3.05 & 3.07 & 2.67 & \textbf{2.50} & 2.64 & 2.49 & 3.07 \\
barley-pp & 7 & 26 & 78 & 0.43 & \textbf{0.43} & 0.45 & 0.58 & 0.55 & 0.60 & 0.60 & 0.66 & 0.54 & 0.51 & 0.53 & 0.44\\
barley2 & 7 & 48 & 126 & 5.67 & 6.80 & 5.65 & 6.11 & 2.79 & 3.50 & 3.17 & 2.97 & 2.24 & 3.02 & \textbf{2.23} & 2.87 \\
celar02 & 10 & 100 & 311 & 61.33 & 227.53 & 114.15 & 41.95 & 4.73 & 8.39 & 6.02 & 6.07 & 3.83 & 5.86 & \textbf{3.80} & 4.40 \\
celar09pp & 7 & 67 & 165 & 3.00 & 10.92 & 4.40 & 4.20 & 0.37 & 0.39 & 0.37 & 0.38 & $\mathbf{<0.1}$ & $\mathbf{<0.1}$ & $\mathbf{<0.1}$ & $\mathbf{<0.1}$ \\
child & 3 & 20 & 30 & $\mathbf{<0.1}$ & $\mathbf{<0.1}$ & $\mathbf{<0.1}$ & $\mathbf{<0.1}$ & $\mathbf{<0.1}$ & $\mathbf{<0.1}$ & $\mathbf{<0.1}$ & $\mathbf{<0.1}$ & $\mathbf{<0.1}$ & $\mathbf{<0.1}$ & $\mathbf{<0.1}$ & $\mathbf{<0.1}$ \\
david & 13 & 87 & 406 & 109.26 & 132.38 & 116.26 & 120.97 & 96.58 & \textbf{60.05} & 106.17 & 110.39 & 78.83 & 78.83 & 82.36 & 97.66 \\
david-pp & 13 & 29 & 191 & 1.92 & 2.47 & 4.54 & 6.31 & 3.95 & 3.15 & 5.86 & 4.53 & 2.30 & 5.25 & 4.09 & \textbf{1.33} \\
eil51.tsp & - & 51 & 140 & TO & TO & TO & TO & TO & TO & TO & TO & TO & TO & TO & TO \\
hailfinder & 4 & 56 & 99 & 1.84 & 4.20 & 2.47 & 2.25 & 0.37 & 0.45 & 0.39 & 0.40 & $\mathbf{<0.1}$ & $\mathbf{<0.1}$ & $\mathbf{<0.1}$ & $\mathbf{<0.1}$ \\
hepar2 & 6 & 70 & 158 & 3.38 & 10.68 & 4.58 & 4.65 & 0.59 & 0.60 & 0.57 & 0.60 & $\mathbf{<0.1}$ & $\mathbf{<0.1}$ & $\mathbf{<0.1}$ & $\mathbf{<0.1}$ \\
huck & 10 & 74 & 301 & 4.34 & 14.82 & 5.37 & 5.59 & 1.58 & 1.60 & 1.58 & 1.70 & $\mathbf{<0.1}$ & $\mathbf{<0.1}$ & $\mathbf{<0.1}$ & $\mathbf{<0.1}$ \\
jean & 9 & 77 & 254 & 4.82 & 17.59 & 7.04 & 6.27 & 1.75 & 2.03 & 1.77 & 1.97 & $\mathbf{<0.1}$ & $\mathbf{<0.1}$ & $\mathbf{<0.1}$ & $\mathbf{<0.1}$ \\
mainuk & 7 & 48 & 198 & 1.16 & 1.98 & 1.57 & 1.52 & 1.60 & 2.01 & 1.78 & 1.79 & $\mathbf{<0.1}$ & $\mathbf{<0.1}$ & $\mathbf{<0.1}$ & $\mathbf{<0.1}$ \\
mildew & 4 & 35 & 80 & 0.45 & 0.72 & 0.50 & 0.53 & 0.29 & 0.31 & 0.29 & 0.31 & $\mathbf{<0.1}$ & $\mathbf{<0.1}$ & $\mathbf{<0.1}$ & $\mathbf{<0.1}$ \\
miles500 & - & 128 & 1170 & TO & TO & TO & TO & TO & TO & TO & TO & TO & TO & TO & TO \\
mulsol.i.5-pp & 31 & 119 & 2556 & TO & TO & TO & TO & TO & TO & TO & TO & \textbf{44.80} & 227.82 & 45.11 & 147.31 \\
myciel3 & 5 & 11 & 20 & $\mathbf{<0.1}$ & $\mathbf{<0.1}$ & $\mathbf{<0.1}$ & $\mathbf{<0.1}$ & $\mathbf{<0.1}$ & $\mathbf{<0.1}$ & $\mathbf{<0.1}$ & $\mathbf{<0.1}$ & $\mathbf{<0.1}$ & $\mathbf{<0.1}$ & $\mathbf{<0.1}$ & $\mathbf{<0.1}$ \\
myciel4 & 10 & 23 & 71 & 16.14 & 11.58 & 13.62 & 17.28 & 20.23 & 11.97 & 21.87 & 15.15 & \textbf{7.11} & 11.99 & 7.11 & 16.05 \\
myciel5 & - & 47 & 236 & TO & TO & TO & TO & TO & TO & TO & TO & TO & TO & TO & TO \\
oesoca & 3 & 39 & 67 & 0.53 & 0.86 & 0.62 & 0.70 & $\mathbf{<0.1}$ & $\mathbf{<0.1}$& $\mathbf{<0.1}$& $\mathbf{<0.1}$& $\mathbf{<0.1}$ & $\mathbf{<0.1}$ & $\mathbf{<0.1}$ & $\mathbf{<0.1}$\\
oesoca+-pp & 11 & 14 & 75 & 0.36 & \textbf{0.26} & 0.31 & 0.46 & 0.48 & 0.57 & 0.48 & 0.59 & 0.51 & 0.41 & 0.50 & 0.53 \\
oesoca+ & 11 & 67 & 208 & TO & TO & TO & TO & 453.84 & 417.94 & \textbf{270.68} & 319.46 & 436.02 & 391.96 & 292.28 & 352.75 \\
oesoca42 & 3 & 42 & 72 & 0.65 & 1.06 & 0.73 & 0.88 & 0.11 & 0.12 & 0.12 & 0.12 & $\mathbf{<0.1}$ & $\mathbf{<0.1}$ & $\mathbf{<0.1}$ & $\mathbf{<0.1}$ \\
pathfinder & 6 & 109 & 211 & 13.36 & 119.24 & 19.56 & 17.66 & 0.60 & 0.65 & 0.60 & 0.63 & $\mathbf{<0.1}$ &$\mathbf{<0.1}$ & $\mathbf{<0.1}$ & $\mathbf{<0.1}$ \\
pathfinder-pp & 6 & 12 & 43 & 0.03 & 0.03 & 0.03 & $\mathbf{<0.1}$ & $\mathbf{<0.1}$ & $\mathbf{<0.1}$ & $\mathbf{<0.1}$ & $\mathbf{<0.1}$ & $\mathbf{<0.1}$ & $\mathbf{<0.1}$ & $\mathbf{<0.1}$ & $\mathbf{<0.1}$ \\
queen5\_5 & - & 25 & 160 & TO & TO & TO & TO & TO & TO & TO & TO & TO & TO & TO & TO \\
queen6\_6 & - & 36 & 290 & TO & TO & TO & TO & TO & TO & TO & TO & TO & TO & TO & TO \\
queen7\_7 & - & 49 & 476 & TO & TO & TO & TO & TO & TO & TO & TO & TO & TO & TO & TO \\
ship-ship-pp & 8 & 30 & 77 & 68.17 & \textbf{53.49} & 60.68 & 68.69 & 53.72 & 55.48 & 58.28 & 58.00 & 61.94 & 63.23 & 62.93 & 62.14 \\
water & 9 & 32 & 123 & 1.06 & 2.34 & 2.54 & 0.95 & 1.26 & 1.44 & 1.84 & 1.51 & 0.71 & 0.97 & \textbf{0.54} & 0.67 \\
win95pts & 8 & 76 & 225 & 6.02 & 16.41 & 8.99 & 7.64 & 1.94 & 2.37 & 2.05 & 2.17 & 2.09 & \textbf{1.82} & 2.09 & 1.92 \\
\bottomrule
\end{tabular}}
\caption{Running times (seconds) on the Berg--Jarvisalo benchmark for our SAT-based configurations.}
\label{tab:berg-results}
\end{table*}

\clearpage
\section{Extended Experimental Evaluation: Pseudo-Random Clustered Graphs}
\label{app:clustered-extended}

This appendix provides additional plots and breakdowns for the pseudo-random clustered benchmark
(Section~\ref{sec:exp-clustered}), complementing the main-paper heatmap view.

\subsection{Example instance}
\label{subsec:app-clustered-example}

\begin{figure}[H]
  \centering
  \includegraphics[width=1.0\linewidth]{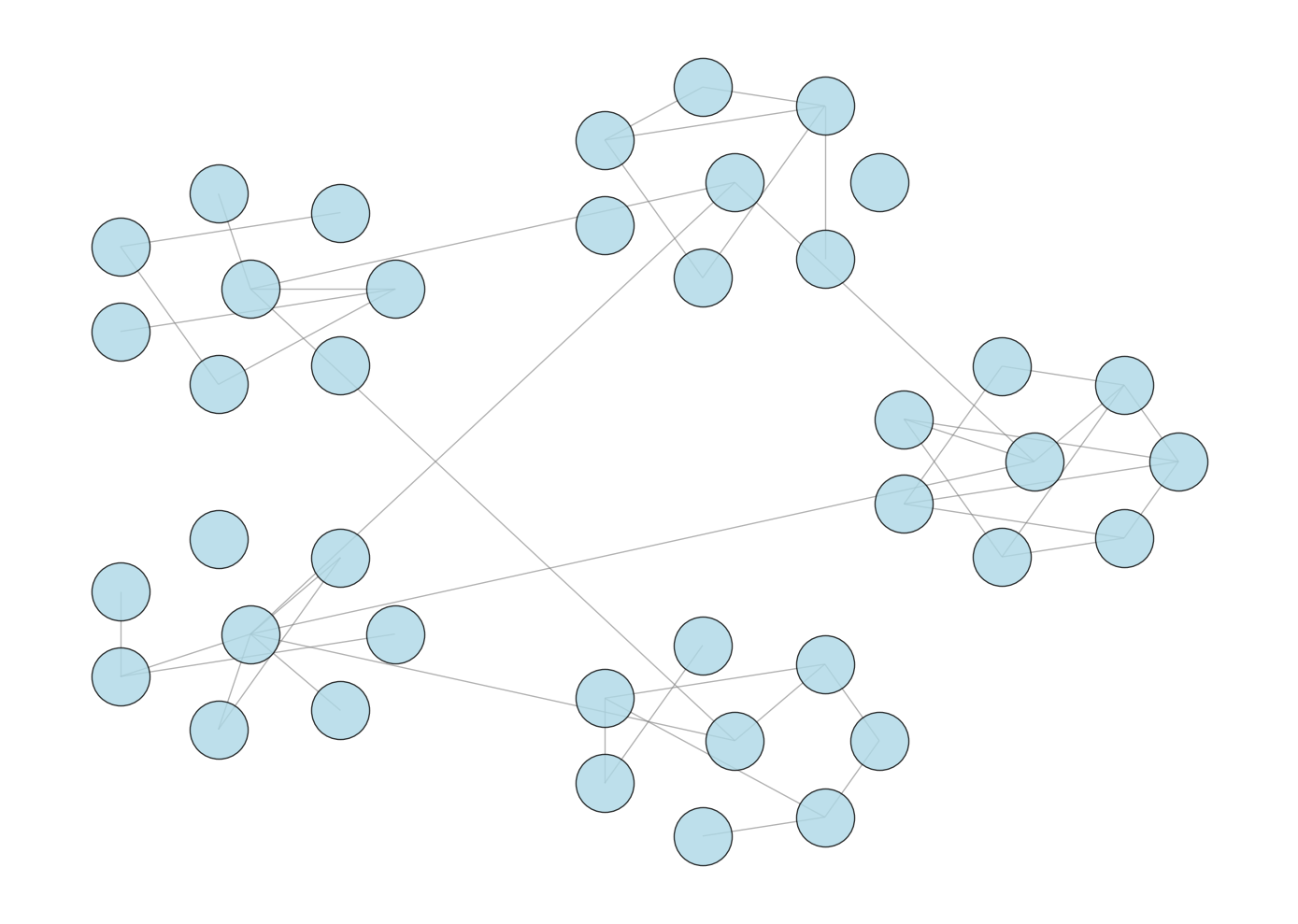}
  \caption{Example pseudo-random clustered graph instance.}
  \label{fig:clustered-example}
\end{figure}

Figure~\ref{fig:clustered-example} illustrates a representative generated instance. Dense intra-cluster
connectivity contrasts with sparse inter-cluster links routed through a small set of gateway vertices,
which is precisely the type of structure that yields many small separators. Each instance is generated by a two-level random process. For given parameters $(k,n)$, the vertex set
is partitioned into $k$ clusters $C_0,\dots,C_{k-1}$ of size $n$, with
\[
C_i=\{ i\cdot n,\; i\cdot n + 1,\; \dots,\; i\cdot n + (n-1)\}.
\]
Within each cluster $C_i$, we sample an Erd\H{o}s--R\'enyi graph $G(n,p_{\text{in}})$ by adding each
intra-cluster edge independently with probability
$p_{\text{in}}=\texttt{cluster\_density}=0.5$.
To control cross-links, we designate a single gateway vertex per cluster, $g_i := in$, and add
inter-cluster edges only between gateways: for each pair $i<j$, we add the edge $\{g_i,g_j\}$
independently with probability
$p_{\text{out}}=\texttt{inter\_cluster\_prob}=0.25$.
No other inter-cluster edges are introduced. Overall, the generated graphs have sizes ranging from 10 to 280 vertices with associated treewidths going from 1 to 13.

\subsection{Scaling with cluster size (fixed $k$)}
\label{subsec:app-clustered-scaling}

\begin{figure}[H]
  \centering
  \includegraphics[width=\linewidth]{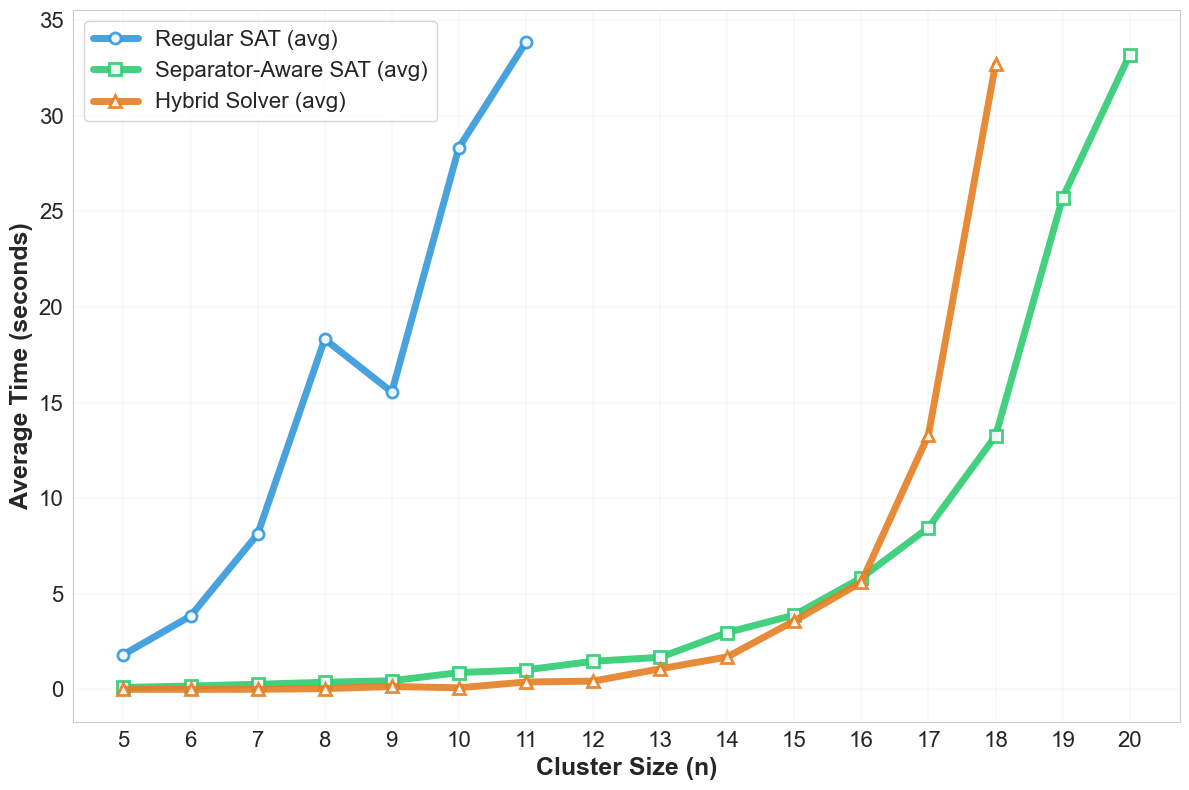}
  \caption{Average runtime versus cluster size $n$ for a fixed number of clusters ($k=11$),
  averaged across atomic solving modes.}
  \label{fig:clustered-scaling}
\end{figure}

To isolate the effect of cluster size, Figure~\ref{fig:clustered-scaling} fixes the number of
clusters ($k=11$) and reports runtime as $n$ increases, averaged
across atomic solving modes.

\noindent The baseline grows sharply and quickly reaches the hard regime.
Both separator-based configurations substantially flatten this trend for small and medium cluster
sizes, consistent with decomposition preventing the atomic solver from facing the full instance.
We can also see that \emph{Separator-aware} remains slightly better than \emph{Hybrid} towards the end of the plot. Note that nevertheless, we can clearly notice the trend of exponential blowup starting from cluster size = 15 thus showing that at some point, both separtor-based enhanced solvers face unavoidably hard atomic subgraphs (the clusters). 

\subsection{Timeouts}
\label{subsec:app-clustered-timeouts}

\begin{figure}[H]
  \centering
  \includegraphics[width=\linewidth]{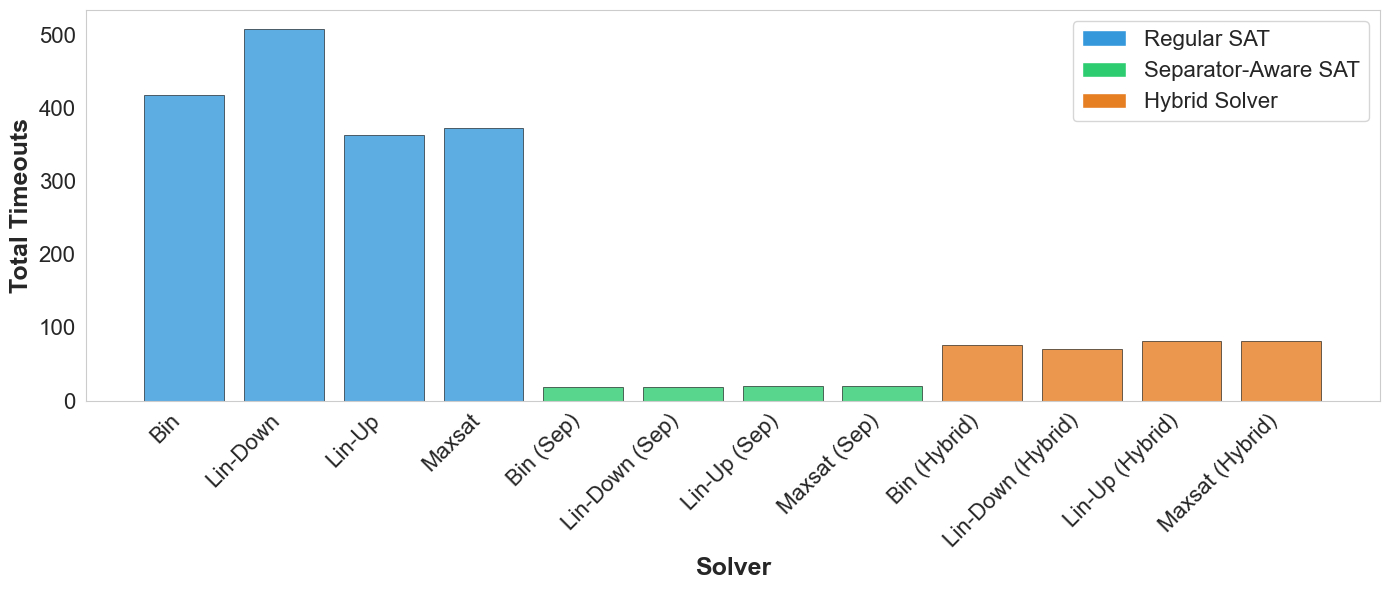}
  \caption{Total number of timeouts per solver configuration and atomic solving mode
  on the pseudo-random clustered benchmark.}
  \label{fig:clustered-timeouts}
\end{figure}

Figure~\ref{fig:clustered-timeouts} summarizes the total number of timeouts across solver
configurations and atomic solving modes.

\noindent The \emph{Regular} baseline times out frequently across all modes.
Both separator-based approaches substantially reduce timeouts, and \emph{Separator-aware} drives them
to near zero on this separator-rich benchmark, consistent with the cross-leaf clause reuse mechanism discussed in Section~\ref{sec:exp-clustered}.

\clearpage
\section{Extended Experimental Evaluation: Random Graphs}
\label{app:random-extended}

We complement the main-paper results with an extended full grid. 

\subsection{More on instance generation}
\label{app:random-gen}

Each instance is sampled from the Erd\H{o}s--R\'enyi model $G(n,p)$: meaning that for a fixed number of vertices
$n$, each potential edge is included independently with probability~$p$. Overall the generated graph have treewidths going from 1 to 7 (with vertices going from 30 to 80).

\subsection{Runtime landscape over $(n,p)$}
\label{app:random-heatmap}

\begin{figure*}[t]
  \centering
  \includegraphics[width=\textwidth]{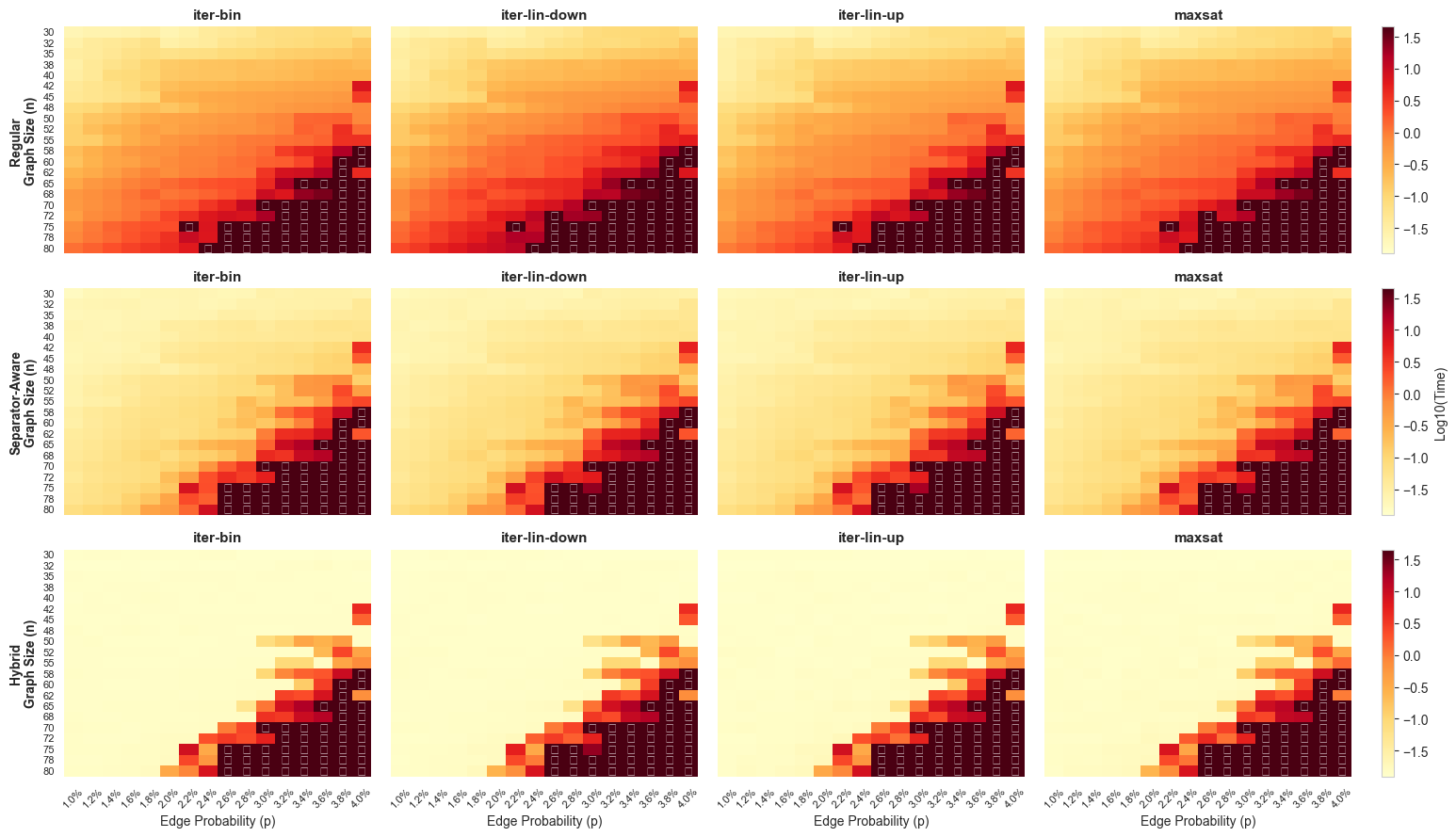}
  \caption{Runtime heatmaps (log$_{10}$ seconds) on Erd\H{o}s--R\'enyi graphs over graph size $n$
  and edge probability $p$, for all solver configurations and modes. White square denotes timeouts.}
  \label{fig:random-heatmap}
\end{figure*}

Figure~\ref{fig:random-heatmap} shows the full runtime landscape over the $(n,p)$ grid on a logarithmic
scale.
For very small $p$, instances are often disconnected and contain many simple separators, so all
configurations tend to solve quickly.
As $p$ increases, the \emph{Regular} baseline enters the hard regime earlier and accumulates
timeouts slightly sooner.
Both separator-based configurations delay this transition in the sparse-to-moderate density range,
but the advantage narrows in the dense regime, where separator structure becomes scarce and all
approaches are increasingly dominated by the underlying SAT/MaxSAT bottleneck near the optimum.

\subsection{Warm-starting via phase hints}
\label{subsec:exp-warmstart}

We complement the main paper’s SAT-layer study by reporting our warm-start experiment. We evaluate warm-starting on \texttt{myciel4.gr} from the Berg--J\"arvisalo benchmark using the
\texttt{iter-lin-down} mode (descending scan). We compute a greedy upper-bound ordering and
translate it into \emph{phase/polarity hints} for precedence variables $\mathit{ord}_{ij}$, without
adding any constraints. We run 5 descents with warm-start enabled and 5 without, recording the mean
cumulative time after each tested width.

\begin{figure}[H]
  \centering
  \includegraphics[width=\linewidth]{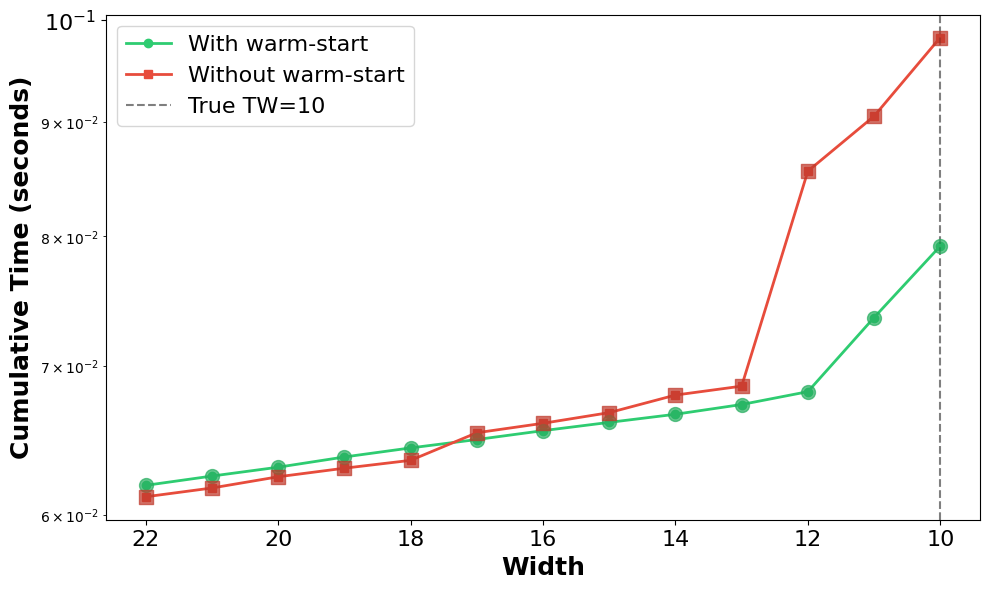}
  \caption{Warm-start impact for \texttt{iter-lin-down} on \texttt{myciel4.gr}.
  Curves show mean cumulative time (5 runs) after testing each width. The dashed line marks the true
  treewidth ($\tw=10$).}
  \label{fig:warmstart-lin-down-myciel4}
\end{figure}

\noindent Figure~\ref{fig:warmstart-lin-down-myciel4} shows a small improvement on this
instance, most visible near the optimum. More broadly, the effect is instance-dependent and usually
modest.

\section{Extended Experimental Evaluation: PACE2017 (Exact Treewidth)}
\label{app:pace-extended}

This appendix provides additional empirical statistics complementing
Section~\ref{sec:exp-pace} on the PACE2017 \emph{regular} instances.
We focus on quantifying the effect of safe-separator decomposition \emph{prior} to invoking the
atomic SAT backend.

\subsection{Reduction ratios}
\begin{figure}[H]
  \centering
  \includegraphics[width=\linewidth]{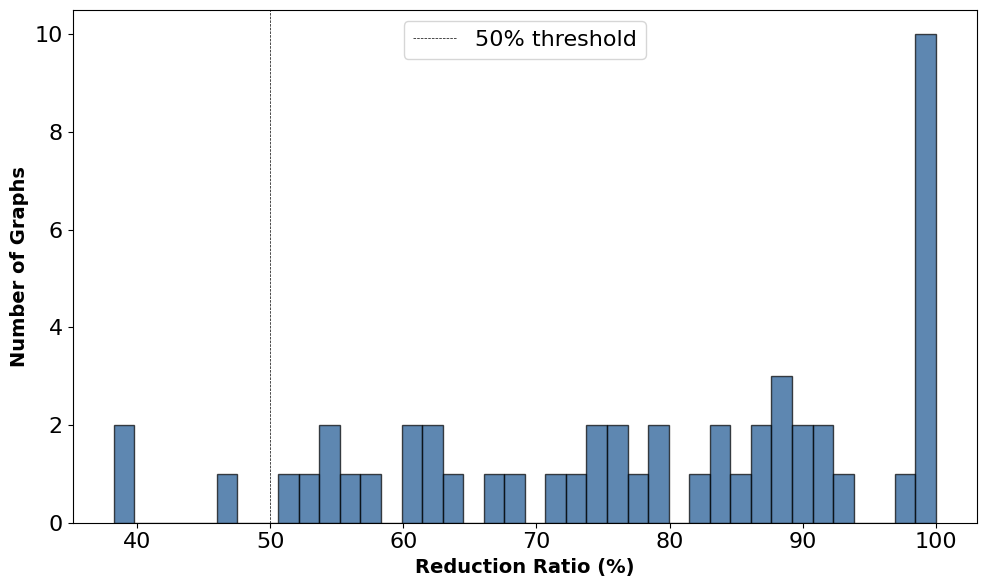}
  \caption{PACE2017: distribution of reduction ratios induced by safe-separator decomposition.
  The dashed vertical line marks a $50\%$ reduction threshold.}
  \label{fig:pace-reduction-ratios}
\end{figure}

Figure~\ref{fig:pace-reduction-ratios} shows the distribution of reduction ratios, which is another view from the main paper to portray the limited decomposability of the PACE2017 instances. A reduction ratio of 100\% means that decomposition achieves no size reduction: no safe separator is found and the largest remaining subproblem has the same size as the original graph.

\subsection{Detected safe-separator types}
Table~\ref{tab:pace-separator-types} summarizes which safe-separator classes are detected during
recursive decomposition on PACE2017 (disconnected components omitted since all considered instances
are connected).

\begin{table}[t]
\centering
\caption{Distribution of detected safe-separator types on the PACE2017 benchmark
(regular instances only).}
\label{tab:pace-separator-types}
\begin{tabular}{lrr}
\toprule
\textbf{Separator type} & \textbf{Count} & \textbf{Proportion} \\
\midrule
Clique separators        & 1214 & 95.4\% \\
Almost-clique separators & 50   & 3.9\%  \\
Articulation points      & 8    & 0.6\%  \\
\bottomrule
\end{tabular}
\end{table}

\noindent
On average, 25.44 safe separators per graph are detected across the benchmark. While this seems consequent, usually these separators are highly unbalanced (far from ideal separators which would split in half each graph instance).

\section{Extended Experimental Evaluation: Bound Heuristics Accuracy}
\label{app:bounds-extended}

This appendix simply provides an additional statistic to  Section~\ref{sec:exp-bounds}.

\subsection{Aggregate accuracy statistics}
\label{appsubsec:exp-bounds-aggregate}

Table~\ref{tab:bounds-summary} summarizes aggregate bound quality on PACE2017.
The greedy upper bound is typically close to optimal, while MMW is looser but still provides a
useful floor. The resulting gap $UB-LB$ is often small enough to substantially restrict the
width interval explored by the atomic iterative SAT/MaxSAT backend, nevertheless as stated, the hardest part usually lies when close to the optimal treewidth, explaining why SAT-based solver even with enhanced upper and lower bounds timeout on the PACE2017 instances.

\begin{table}[H]
\caption{Accuracy of upper and lower bound heuristics on PACE2017 instances.}
\label{tab:bounds-summary}
\centering
\begin{tabular}{lrr}
\toprule
\textbf{Metric} & \textbf{Upper Bound} & \textbf{Lower Bound} \\
\midrule
Mean absolute error & 2.46 & 5.93 \\
Max absolute error  & 12.0 & 26.0 \\
Exact matches (\%) & 21.5 & 10.0 \\
\midrule
\textbf{Gap statistics} & \multicolumn{2}{c}{ } \\
\midrule
Mean gap $UB-LB$ & \multicolumn{2}{c}{8.38} \\
Median gap       & \multicolumn{2}{c}{6} \\
Max gap          & \multicolumn{2}{c}{35} \\
Instances with gap $=0$ & \multicolumn{2}{c}{20} \\
Instances with gap $\le 5$ & \multicolumn{2}{c}{77} \\
\bottomrule
\end{tabular}
\end{table}

\clearpage

\section{Implementation Details}
This appendix gives implementation details orthogonal to the main theoretical development, to support reproducibility and clarify core routines in pseudo-code form. The full implementation is available on GitHub: \texttt{https://github.com/NaykinYT/Treewidth}.\\

\noindent Additional documentation is provided in \texttt{src/README}, while experiment and testing instructions are in \texttt{benchmark/README} and \texttt{tests/README} (see also the top-level \texttt{README}).

\subsection{Detection of disconnected components}
\label{sec:impl-disconnected}

Our implementation uses \textsc{NetworkX}~\cite{hagberg2008networkx} to
test connectivity and enumerate components, returning relabeled component subgraphs together with
vertex maps old$\leftrightarrow$new for later decoding and merging.

\paperparagraph{Complexity.} $\mathcal{O}(|V|+|E|)$.

\paperparagraph{Code.}
\texttt{src/separator/safe/finders/disconnected.py}.

\subsection{Detection of articulation points}
\label{sec:impl-articulation}

Our implementation relies on
\textsc{NetworkX}~\cite{hagberg2008networkx} to enumerate articulation points. For each articulation point $a$, we form the
singleton separator $S=\{a\}$, remove $S$ from the graph, and compute the connected components of
$G\setminus S$ and keep track of them. Each
candidate is assigned a score (using Algorithm ~\ref{sec:impl-sep-score}) sorted by increasing score.

\paperparagraph{Complexity.} $\mathcal{O}(|V|(|V|+|E|))$
\paperparagraph{Code.}
\texttt{src/separator/safe/finders/articulation.py}.

\subsection{Detection of clique separators}
\label{sec:impl-clique}

Clique separators are generated by enumerating cliques $S$ of size at least $2$ up to a fixed limit
$k_{\max}$ and testing whether removing $S$ disconnects the graph using \textsc{NetworkX}~\cite{hagberg2008networkx}. We keep $k_{\max}$ small in
practice, typically $k_{\max}=3$, since the number of clique candidates grows rapidly with
$k_{\max}$. Enumeration proceeds by increasing clique size, so it can be stopped as soon as a clique
larger than $k_{\max}$ is reached. For each candidate clique $S$, we compute the connected components
of $G\setminus S$ using a subgraph view, discard candidates yielding fewer than two components, and just like for articulation nodes separator, we assign it a score using  Algorithm ~\ref{sec:impl-sep-score}. All detected clique separators are
returned sorted by increasing score.

\paperparagraph{Complexity.}
Let $n=|V|$ and $m=|E|$, and let $C_{\le k_{\max}}$ be the number of enumerated cliques of size at
most $k_{\max}$. Each candidate requires a connected-components computation on $G\setminus S$, which
takes $\mathcal{O}(n+m)$ time, so the overall worst-case runtime is
$\mathcal{O}(C_{\le k_{\max}}\cdot(n+m))$. In the worst case $C_{\le k_{\max}}=\mathcal{O}(n^{k_{\max}})$,
yielding $\mathcal{O}(n^{k_{\max}}(n+m))$ for fixed $k_{\max}$.

\paperparagraph{Code.}
\texttt{src/separator/safe/finders/clique\_separator.py}.

\subsection{Detection of almost-clique separators}
\label{sec:impl-almost-clique}

We follow Tamaki's $A(H,G)$ heuristic~\cite{Tamaki2021AlmostClique}: we compute a chordal
completion $H$ of $G$ (via Maximum Cardinality Search (MCS) followed by triangulation), build a clique tree of $H$, and use
intersections of adjacent maximal cliques as candidate separators. Each candidate $S$ is
retained only if $2\le |S|\le k_{\max}$, $S$ is a minimal separator of $G$ (i.e., $G\setminus S$ is disconnected, and no proper subset $S'\subsetneq S$ already disconnects $G$), and $S$ is an almost-clique in $G$.

\paperparagraph{Complexity.}
MCS is $\mathcal{O}(n+m)$, while our triangulation step is
$\mathcal{O}\!\left(\sum_{v\in V} (d^+_{\pi}(v))^2\right)$ (worst-case $\mathcal{O}(n^3)$);
clique-tree construction by pairwise clique intersections is worst-case $\mathcal{O}(c^2 n)$.
Candidate validation is exponential only in $|S|$ and thus bounded by a small constant for
fixed $k_{\max}$.

\paperparagraph{Code.}
\texttt{src/separator/safe/finders/almost\_clique.py}.

\subsection{Decoding SAT models into tree-decompositions}
\label{sec:impl-decode-td}

This subsection explains how we reconstruct an explicit tree decomposition from a satisfying SAT
assignment of the iterative Berg--J{\"a}rvisalo encoding (following Bodlaender and Koster ~\cite{BodlaenderKoster2010UpperBounds}).
A model $M$ (result of the assignements of the SAT/MaxSAT solver) fixes (i) a total order via $\textit{ord}_{ij}$ and (ii) an (oriented) triangulation via
the $O_{ij}$ variables.
We extract an elimination ordering $\pi$, normalize the $O$-edges to follow $\pi$, derive the later-neighbor
sets $N^+(u)$, and then build bags and tree edges using the standard elimination-order construction.

\paperparagraph{Bag construction and connectivity rule.}
Given $\pi=(v_1,\dots,v_n)$ and later-neighbor sets $N^+(u)$ (neighbors of $u$ that appear later in $\pi$
in the triangulated graph), we create one bag per vertex,
\[
  X_u := \{u\}\cup N^+(u).
\]
To connect bags into a tree, we use the parent-by-earliest-later-neighbor rule:
if $N^+(u)\neq\emptyset$, let
\[
  p(u) := \arg\min_{v\in N^+(u)} \mathrm{pos}(v),
\]
i.e., the earliest vertex of $N^+(u)$ under $\pi$, and add the tree edge $(X_u, X_{p(u)})$.
If $N^+(u)=\emptyset$, $X_u$ becomes a root bag (or is attached to a designated root, which is
irrelevant for width).
This attachment ensures $N^+(u)\subseteq X_{p(u)}$, and yields the running-intersection property (from Definition \ref{def:treewidth-td}).

\paperparagraph{Complexity.}
Let $n=|V|$ and let $|O|$ be the number of extracted oriented triangulation edges.
Computing $\mathrm{pos}$ takes $\mathcal{O}(n)$.
Extracting and normalizing positive $O_{ij}$ literals takes $\mathcal{O}(|O|)$.
Building all $N^+(u)$ takes $\mathcal{O}(|O|)$.
Constructing bags and parent pointers is $\mathcal{O}\!\left(n+\sum_u |N^+(u)|\right)=\mathcal{O}(n+|O|)$,
since each oriented edge contributes once to exactly one $N^+(u)$.
Thus the overall decoding runs in $\mathcal{O}(n+|O|)$, and in the worst case $|O|=\Theta(n^2)$.

\paperparagraph{Code.}\mbox{}\\
SAT-side decoding and $O$-edge extraction: \\
\noindent \texttt{src/sat\_engine/encoding/iterative.py}.\\
Decomposition construction utilities: \\ \noindent \texttt{src/core/decomposition.py}.

\begin{algorithm}[H]
\caption{\textsc{DecodeTD}$(M)$ (SAT model $\to$ tree-decomposition)}
\label{alg:impl-decode-td}
\begin{algorithmic}[1]
\Require Satisfying assignment $M$ for $F_{\text{iter}}(G)$ on $n$ vertices
\Ensure Tree-decomposition $\mathcal{D}=(\mathrm{bags},\mathrm{td\_edges},\mathrm{width})$
\State $\pi \gets$ \Call{ExtractOrderingFromModel}{$M$} \Comment{$\pi=(v_1,\dots,v_n)$}
\State $\mathrm{pos}[v_i]\gets i$ for $i=1,\dots,n$
\State $O\gets\emptyset$
\ForAll{positive literals $O_{ij}$ in $M$}
  \If{$\mathrm{pos}[i] < \mathrm{pos}[j]$} \State $O \gets O \cup \{(i,j)\}$
  \Else \State $O \gets O \cup \{(j,i)\}$ \Comment{normalize direction to follow $\pi$}
  \EndIf
\EndFor
\State $N^+(u)\gets\emptyset$ for all $u\in V$
\ForAll{$(u,v)\in O$}
  \State $N^+(u)\gets N^+(u)\cup\{v\}$
\EndFor
\State $\mathrm{bags}\gets\emptyset$, \ $\mathrm{td\_edges}\gets[\ ]$, \ $\mathrm{width}\gets 0$
\For{$i=1$ to $n$} \Comment{iterate in elimination order}
  \State $u \gets v_i$
  \State $X_u \gets \{u\}\cup N^+(u)$
  \State $\mathrm{bags}[u]\gets X_u$
  \State $\mathrm{width}\gets \max(\mathrm{width},\,|X_u|-1)$
  \If{$N^+(u)\neq\emptyset$}
    \State $p(u)\gets \arg\min_{v\in N^+(u)} \mathrm{pos}[v]$
    \State $\mathrm{td\_edges}.\mathrm{append}\bigl((u,\,p(u))\bigr)$
  \EndIf
\EndFor
\State \Return $(\mathrm{bags},\mathrm{td\_edges},\mathrm{width})$
\end{algorithmic}
\end{algorithm}

\subsection{Bounds-aware iterative SAT and MaxSAT}
\label{sec:impl-bounds-sat}

Bounds-aware iterative SAT and MaxSAT are implemented as an extension of the baseline
Berg--J\"arvisalo backend described in the main text (Section~\ref{sec:baseline-sat}). Iterative
probing strategies operate only on the current interval $[LB,UB]$, and the MaxSAT export restricts
the optimization to thresholds within this interval (as described in the main paper).

\paperparagraph{Code.}
\texttt{src/sat\_engine/encoding/} and \\ \texttt{src/sat\_engine/algorithm/}.

\subsection{Upper- and lower-bound heuristics}
\label{sec:impl-bounds}

\paperparagraph{Lower bound.}
Our lower bound is the minor-min-width heuristic (MMW) from Bodlaender and Koster ~\cite{BodlaenderKoster2011LowerBounds}, which repeatedly contracts edges
to form a sequence of minors (Definition~\ref{def:minor}).
By minor-monotonicity, contracting edges cannot increase treewidth, hence any width witnessed
along the contraction process yields a valid lower bound on $\tw(G)$.
MMW maintains a working graph $H$ (initially $G$), repeatedly selects a minimum-degree vertex
$v$, updates the bound with $\deg_H(v)$, and then contracts $v$ into a minimum-degree neighbor
$u \in N_H(v)$ (removing $v$ and connecting $u$ to all neighbors of $v$).

\begin{algorithm}[H]
\caption{MMW lower bound}
\label{alg:mmw}
\begin{algorithmic}[1]
\Require Graph $G=(V,E)$
\State $H \gets G$ \Comment{working copy}
\State $LB \gets 0$
\While{$V(H)\neq \emptyset$}
  \State $v \gets \arg\min_{x\in V(H)} \deg_H(x)$
  \State $LB \gets \max(LB,\deg_H(v))$
  \If{$\deg_H(v)=0$}
    \State remove $v$ from $H$
    \State \textbf{continue}
  \EndIf
  \State $u \gets \arg\min_{x\in N_H(v)} \deg_H(x)$
  \State contract edge $(u,v)$ in $H$ \Comment{merge $v$ into $u$}
\EndWhile
\State \Return $LB$
\end{algorithmic}
\end{algorithm}

\paperparagraph{Upper bound: greedy elimination (min-fill / min-degree / hybrid).}
For the upper bound we construct an elimination ordering $\prec$ and take its elimination width
(Definition~\ref{def:elimination-width}), which always satisfies $\tw(G)\le w(\prec)$.
We use standard greedy elimination heuristics from Bodlaender and Koster ~\cite{BodlaenderKoster2010UpperBounds} such as \emph{min-degree} and \emph{min-fill}, as well
as a practical \emph{hybrid} variant.
All variants repeatedly eliminate a vertex $v$ from a working graph $H$:
we record the set of \emph{later neighbors} $N_H(v)$ (the neighbors still present at elimination time),
update the running maximum width with $|N_H(v)|$, add fill edges to make $N_H(v)$ a clique,
and finally remove $v$.
The produced ordering is used both as an upper bound and as a warm-start candidate for SAT
(Section~\ref{sec:impl-warmstart}). \\

\begin{algorithm}[H]
\caption{Greedy upper bound via elimination ordering}
\label{alg:greedy-ub}
\begin{algorithmic}[1]
\Require Graph $G=(V,E)$, method $\in\{\texttt{min-degree},\texttt{min-fill},\texttt{hybrid}\}$
\State $H \gets G$ \Comment{working copy}
\State $\prec \gets [\ ]$, \; $later \gets$ empty map, \; $UB \gets 0$
\State initialize heuristic scores (degree and optionally fill-scores)
\While{$V(H)\neq \emptyset$}
  \State select $v\in V(H)$ according to method
  \State $N \gets N_H(v)$
  \State append $v$ to $\prec$; \quad $later[v] \gets N$
  \State $UB \gets \max(UB, |N|)$
  \ForAll{unordered pairs $\{x,y\}\subseteq N$}
    \State add fill edge $\{x,y\}$ to $H$ if missing
  \EndFor
  \State remove $v$ from $H$
  \State update heuristic scores locally (for affected vertices)
\EndWhile
\State $\mathcal{D} \gets$ \Call{BuildTDFromOrdering}{$\prec$, $later$}
\State \Return $UB$ and $\{\texttt{ordering}=\prec,\ \texttt{decomposition}=\mathcal{D}\}$
\end{algorithmic}
\end{algorithm}

\noindent The difference between the heuristics lies only in how the next vertex is chosen:
\emph{min-degree} picks a minimum-degree vertex;
\emph{min-fill} picks a vertex minimizing the number of missing edges inside its neighborhood
(ties broken by degree);
\emph{hybrid} evaluates fill-scores only for vertices up to a fixed degree threshold $\tau$
and otherwise falls back to min-degree (a common practical trick to control fill-score overhead).
Algorithm~\ref{alg:greedy-ub} summarizes the shared control flow.
\paperparagraph{Complexity.} Throughout the experiments we used the min-fill heuristic for the hybrid solver. 
Both bounds (lower and upper bound variants) run in polynomial time.

\paperparagraph{Code.}\mbox{}\\
\noindent\texttt{src/approximator/lowerbound/lower\_bounds.py},\\
\texttt{src/approximator/upperbound/upper\_bound.py},\\
\texttt{src/approximator/utils.py},\\

\subsection{Warm-start hints to the SAT solver}
\label{sec:impl-warmstart}

Whenever the greedy upper-bound heuristic returns an elimination ordering $\prec$,
we use it to \emph{warm-start} the exact solver by providing \emph{phase/polarity hints}
rather than adding constraints.
This preserves exactness: the SAT/MaxSAT instance is unchanged, and the solver remains free
to flip any variable if needed. Let $\pi:V\to\{1,\dots,n\}$ map each vertex to its position in $\prec$.
For each ordering variable $\textit{ord}_{ij}$ (encoding whether $v_i \prec v_j$),
we set the preferred polarity to match the heuristic order:
\[
  \textit{ord}_{ij}^{\star} \;=\;
  \begin{cases}
    1 & \text{if } \pi(v_i) < \pi(v_j),\\
    0 & \text{otherwise.}
  \end{cases}
\]

\subsection{Merging tree-decompositions}
\label{sec:impl-merge}

Both \emph{HybridSolver} (Section~\ref{sec:hybrid}) and the separator-aware decoding
(Section~\ref{sec:sep-aware-encoding}) produce several child tree-decompositions that must be
combined into a single tree-decomposition for the parent instance.
At an internal decomposition step with separator $S$, each child subinstance contains $S$
(and after separator augmentation, $S$ is a clique), so the decompositions can be merged by
introducing a \emph{hub} bag containing $S$ and connecting every child decomposition to this hub,
following the standard separator-based recombination of tree decompositions~\cite{BodlaenderKoster2010UpperBounds}.

\noindent Concretely, given child decompositions $\mathcal{D}_1,\dots,\mathcal{D}_r$, we build a new
decomposition by (i) creating a fresh hub bag $X_{\text{hub}} := S$, (ii) copying all child bags
and tree edges into a global structure while relabeling their bag identifiers to avoid collisions,
and (iii) attaching each child decomposition to the hub through a bag that contains $S$ (which
exists by construction). To keep hub attachments small, we choose the \emph{smallest} such bag in
each child. The width of the merged decomposition is
$\max\bigl(\max_i \mathrm{width}(\mathcal{D}_i),\,|S|-1\bigr)$.

\paperparagraph{Complexity.}
Let $B$ and $E$ denote the total number of bags and tree edges across all child decompositions.
Merging consists of copying and relabeling all bags and edges, plus adding $r$ hub-attachment edges.
If we account for copying bag contents, the runtime is linear in the size of the resulting decomposition:
\[
  \mathcal{O}\!\left(B + E + \sum_{X \in \mathrm{bags}} |X|\right),
\]
where the sum ranges over all bags $X$ in the \emph{merged} decomposition (including the hub), and $|X|$
denotes the number of vertices in bag $X$.

\begin{algorithm}[H]
\caption{\textsc{MergeDecompositions}$(S,\{(V_i,\mathcal{D}_i)\}_{i=1}^r)$}
\label{alg:merge-decompositions}
\begin{algorithmic}[1]
\Require Separator $S$; for each child $i$: vertex set $V_i$ and decomposition
$\mathcal{D}_i=(\mathrm{bags}_i,\mathrm{edges}_i,\mathrm{width}_i)$
\Ensure Merged decomposition $\mathcal{D}=(\mathrm{bags},\mathrm{edges},\mathrm{width})$
\State $\mathrm{bags} \gets \{\,0 \mapsto S\,\}$ \Comment{hub bag id $0$}
\State $\mathrm{edges} \gets [\ ]$
\State $\textit{offset} \gets 1$, \quad $\textit{maxChildWidth} \gets \max_i \mathrm{width}_i$
\For{$i=1$ to $r$}
    \ForAll{$(b,X)\in \mathrm{bags}_i$}
        \State $\mathrm{bags}[b+\textit{offset}] \gets X$
    \EndFor
    \ForAll{$(u,v)\in \mathrm{edges}_i$}
        \State $\mathrm{edges}.\mathrm{append}((u+\textit{offset},\,v+\textit{offset}))$
    \EndFor
    \State $\textit{attach} \gets \arg\min_{b\in \mathrm{bags}_i,\ S\subseteq \mathrm{bags}_i[b]} |\mathrm{bags}_i[b]|$
    \State $\mathrm{edges}.\mathrm{append}((0,\,\textit{attach}+\textit{offset}))$
    \State $\textit{offset} \gets \textit{offset} + |\mathrm{bags}_i|$
\EndFor
\State $\mathrm{width} \gets \max(\textit{maxChildWidth},\,|\mathrm{bags}[0]|-1)$
\State \Return $(\mathrm{bags},\mathrm{edges},\mathrm{width})$
\end{algorithmic}
\end{algorithm}

\paperparagraph{Code.}
\texttt{src/core/decomposition.py}

\subsection{Separator scoring heuristic}
\label{sec:impl-sep-score}

Several separator detectors (notably clique and almost-clique enumeration) may produce many
candidates.  To decide which separator to apply first, we attach to each candidate separator
$S$ a simple \emph{lexicographic score} computed from the sizes of the connected components of
$G \setminus S$. Let $C_1,\dots,C_r$ be the connected components of $G \setminus S$ and let
$s_i = |C_i|$ denote their sizes.  Our implementation defines
\[
  \mathrm{score}(S)
  \;=\;
  \Bigl(
    \max_{i} s_i,\;
    \frac{\max_i s_i}{\sum_i s_i},\;
    r
  \Bigr),
\]
and ranks separators by increasing score (lexicographic order), i.e., \emph{lower is better}.
Equivalently, we first minimize the size of the \emph{largest} resulting component, then break
ties by preferring more balanced splits (small $\max_i s_i / \sum_i s_i$), and finally break
ties by preferring fewer components (smaller $r$) to reduce recursion/merging overhead.

\paperparagraph{Rationale.}
The dominant cost in a decomposition-based exact solver is typically the hardest subinstance.
Thus, the primary objective is to shrink the largest child graph after splitting: if a separator
splits a graph on $n$ vertices into two roughly equal parts of size $n/2$ (ignoring the small
separator itself), then a conceptual ``ordering search space'' viewpoint illustrates the gain:
a monolithic elimination-ordering search ranges over roughly $n!$ orderings, whereas solving two
independent halves has cost on the order of $2\cdot (n/2)!$.  While our SAT/MaxSAT backend does not literally enumerate orderings, its
worst-case behavior remains exponential in the number of vertices of the hardest subinstance,
so separators that split ``near the middle'' are strongly preferred.
The secondary ratio term makes this preference robust across separators of different sizes, and
the final component-count tie-breaker avoids producing an excessive number of tiny subproblems
when two candidates already yield the same largest-part size.

\subsection{Baseline Berg--J{\"a}rvisalo SAT/MaxSAT encoding}
\label{sec:impl-bj-baseline}

The baseline Berg--J{\"a}rvisalo iterative encoding and the corresponding exact solving modes are
presented in the main text (Section~\ref{sec:baseline-sat}); we therefore only point to the
reference implementation.

\paperparagraph{Code (encoding).}\mbox{}\\
\noindent\texttt{src/sat\_engine/encoding/variables.py},\\
\texttt{src/sat\_engine/encoding/base.py},\\
\texttt{src/sat\_engine/encoding/cardinality.py},\\
\texttt{src/sat\_engine/encoding/iterative.py},\\
\texttt{src/sat\_engine/encoding/cnf.py}.

\paperparagraph{Code (solving modes).}\mbox{}\\
\noindent\texttt{src/sat\_engine/algorithms/iter\_lin.py},\\
\texttt{src/sat\_engine/algorithms/iter\_bin.py},\\
\texttt{src/sat\_engine/algorithms/maxsat.py}.

\paperparagraph{Code (warm-start hints).}\mbox{}\\
\texttt{src/sat\_engine/algorithms/warmstart.py}.

\subsection{HybridSolver}
\label{sec:impl-hybrid}

The \textsc{HybridSolver} recursion and control flow are already presented in the main text
(Section~\ref{sec:hybrid}); we therefore only point to the reference implementation.

\paperparagraph{Code.}\mbox{}\\
\texttt{src/hybrid\_engine/hybrid\_solver.py}.

\subsection{Separator-aware SAT encoding}
\label{sec:impl-separator-aware}

The separator-aware iterative encoding and its solver workflows are already described in the main
text (Section~\ref{sec:sep-aware-encoding}); we therefore only point to the reference
implementation.

\paperparagraph{Code (encoding).}\mbox{}\\
\texttt{src/sat\_engine/encoding/iterativeSeparatorAware.py}. \\
\paperparagraph{Code (solving modes).}\mbox{}\\
\texttt{src/sat\_engine/algorithm/iter\_bin\_separator\_aware.py}. \\
\texttt{src/sat\_engine/algorithm/iter\_lin\_separator\_aware.py}. \\
\texttt{src/sat\_engine/algorithm/maxsat\_separator\_aware.py}.

\end{document}